\documentclass{IEEEtran}
\usepackage{cite}
\usepackage{amsmath,amssymb,amsfonts}
\usepackage{graphicx}
\usepackage{textcomp,nicefrac}
\usepackage[bookmarks=false]{hyperref}
\usepackage{url}

\def\BibTeX{{\rm B\kern-.05em{\sc i\kern-.025em b}\kern-.08em
T\kern-.1667em\lower.7ex\hbox{E}\kern-.125emX}}
\begin{document}
\title{A study on energy resolution of CANDLES detector}
\author{B. T. Khai,~\IEEEmembership{Member,~IEEE,} 
	S. Ajimura,~\IEEEmembership{Member,~IEEE,} 
	W. M. Chan,
	K. Fushimi,
	R. Hazama,
	H. Hiraoka,
	T. Iida, 
	K. Kanagawa, 
	H. Kino,
	T. Kishimoto,
	T. Maeda, 
	K. Nakajima,
	M. Nomachi,~\IEEEmembership{Senior Member,~IEEE,} 
	I. Ogawa,\\
	T. Ohata,
	K. Suzuki, 
	Y. Takemoto,
	Y. Takihira,
	Y. Tamagawa,
	M. Tozawa,
	M. Tsuzuki, 
	S. Umehara,\\
	and S. Yoshida
	\thanks{
		Manuscript received MM DD, YYYY; accepted MM DD, YYYY. 
		Date of publication: MM DD, YYYY; date of current version, MM DD, YYYY.
		This work was supported by JSPS/MEXT KAKENHI Grant Number 19H05804, 19H05809, 26104003, 16H00870, 24224007, and 26105513.
		This work was supported by the research project of Research Center for Nuclear Physics~(RCNP), Osaka University.
		This work was also supported by the joint research program of the Institute of Cosmic Ray Research~(ICRR), the University of Tokyo.
		The Kamioka Mining and Smelting Company has provided service for activities in the mine.
	}
	\thanks{B. T. Khai, W. M. Chan, K. Kanagawa, H. Kino, T. Maeda, T. Ohata, M. Tsuzuki, and S. Yoshida are with the Graduate School of Science, Osaka University, Toyonaka, Osaka 560-0043, Japan.}
	\thanks{S. Ajimura, T. Kishimoto, M. Nomachi, Y. Takemoto, Y. Takihira, and S. Umehara are with the Research Center for Nuclear Physics, Osaka University, Ibaraki, Osaka 567-0047, Japan.}
	\thanks{K. Fushimi is with the Faculty of Integrated Arts and Science, University of Tokushima, Tokushima 770-8502, Japan.}
	\thanks{R. Hazama is with the Faculty of Human Environment, Osaka Sangyo University, Daito, Osaka 574-8530, Japan.}
	\thanks{T. Iida is with the Faculty of Pure and Applied Sciences, University of Tsukuba, Ibaraki 305-8571, Japan.}
	\thanks{H. Hiraoka, K. Nakajima, I. Ogawa, Y. Tamagawa, and M. Tozawa are with the Graduate School of Engineering, University of Fukui, Fukui 910-8507, Japan.}
	\thanks{K. Suzuki is with the Wakasa Wan Energy Research Center, 64-52-1 Nagatani, Tsuruga, Fukui 914-0192, Japan.}
}

\maketitle

\begin{abstract}
In a neutrinoless double-beta decay ($0\nu\beta\beta$) experiment, energy resolution is important to distinguish between $0\nu\beta\beta$ and background events.
CAlcium fluoride for studies of Neutrino and Dark matters by Low Energy Spectrometer (CANDLES) discerns the $0\nu\beta\beta$ of $^{48}$Ca using a CaF$_2$ scintillator as the detector and source. 
Photomultiplier tubes (PMTs) collect scintillation photons.
At the Q-value of $^{48}$Ca, the current energy resolution (2.6$\%$) exceeds the ideal statistical fluctuation of the number of photoelectrons (1.6$\%$). 
Because of CaF$_2$'s long decay constant of 1000~ns, a signal integration within 4000~ns is used to calculate the energy.
The baseline fluctuation ($\sigma_{\rm baseline}$) is accumulated in the signal integration, thus degrading the energy resolution.
This paper studies $\sigma_{\rm baseline}$ in the CANDLES detector, which severely degrades the resolution by 1$\%$ at the Q-value of $^{48}$Ca.
To avoid $\sigma_{\rm baseline}$, photon counting can be used to obtain the number of photoelectrons in each PMT; however, a significant photoelectron signal overlapping probability in each PMT causes missing photoelectrons in counting and reduces the energy resolution.
``Partial photon counting” reduces $\sigma_{\rm baseline}$ and minimizes photoelectron loss.
We obtain improved energy resolutions of 4.5--4.0$\%$ at 1460.8~keV ($\gamma$-ray of $^{40}$K), and 3.3--2.9$\%$ at 2614.5~keV ($\gamma$-ray of $^{208}$Tl).
The energy resolution at the Q-value is estimated to be improved from 2.6$\%$ to 2.2$\%$, and the detector sensitivity for the $0\nu\beta\beta$ half-life of $^{48}$Ca can be improved by 1.09 times. 
\end{abstract}

\begin{IEEEkeywords}
	CaF$_{2}$, photon counting, energy resolution
\end{IEEEkeywords}

\newpage
\section{Introduction}
\label{sec:introduce}
\subsection{Double Beta Decay}
\label{ssec:dbd}
\IEEEPARstart{D}{ouble}-beta decay (DBD) is a transition between two isobaric nuclei (A, Z) and (A, Z + 2) with two decay modes.
For a two-neutrino DBD ($2\nu\beta\beta$) mode, two electrons and two electron-type anti-neutrinos are emitted. 
However, an alternative decay mode can occur without anti-neutrino emission, and this is called neutrinoless DBD ($0\nu\beta\beta$).
The $0\nu\beta\beta$ mode is forbidden in the Standard Model (SM) of particle physics due to its violation of lepton number conservation, but it can probe new physics beyond the SM \cite{Rev0nbb}. 
The discovery of neutrino oscillation indicates that a neutrino has non-zero mass \cite{PhysRevLett.86.5651, PhysRevLett.89.011301}; however, the absolute neutrino mass is unknown.
$0\nu\beta\beta$ only occurs if a neutrino is a massive Majorana particle {\cite{Rev0nbb}}, and the effective neutrino mass can be determined by its decay half-life {\cite{Rev0nbb}}. 
The remaining questions related to the neutrino mass and whether neutrinos are Majorana or Dirac fermions are attracting the interest of physicists, and the $0\nu\beta\beta$ experiment is a useful tool for these purposes.
The $2\nu\beta\beta$ mode has been experimentally observed (e.g. the $2\nu\beta\beta$ half-lives of different DBD isotopes are summarized in \cite{2vbb}), but the $0\nu\beta\beta$ mode has not been observed yet.
$^{48}$Ca has the highest DBD Q-value ($Q_{\beta\beta}$~=~4272~keV), but its natural abundance is very low (0.187$\%$).
The highest $Q_{\beta\beta}$ gives a large phase-space factor to enhance the DBD rate and has the least contribution from the natural background.
\subsection{CANDLES experiment}
\label{ssec:candles}
The CAlcium fluoride for studies of Neutrino and Dark matters by Low Energy Spectrometer (CANDLES) experiment aims to observe $0\nu\beta\beta$ from $^{48}$Ca.
The experiment is very challenging owing to the extremely low decay rate of $0\nu\beta\beta$ from $^{48}$Ca ($\rm{T_{1/2}^{0\nu}}>$5.6$\times$10$^{22}$~years with 90$\%$~confidence~level \cite{ajimura2020low}).
To observe $0\nu\beta\beta$, a large amount of the source and a low background measurement are required.
The current CANDLES III experiment is constructed in the Kamioka Underground Observatory (2700~m water equivalent depth) to reduce the cosmic-ray background \cite{Suzuki2012}.
We set up an experiment with 96 CaF$_2$ (un-doped, non-enriched) $10\times10\times10$~cm$^3$ crystals used as both the detector and source.
All crystals are submerged inside a 2~m$^3$ vessel filled with liquid scintillator (LS).
The scintillation decay constants of CaF$_2$ and LS are 1000~ns and 10~ns, respectively, and the LS is used as a 4$\pi$ active shielding.
Scintillation photons are collected by 62~photomultiplier tubes (PMTs) surrounding the vessel, of which 12 are 10-inch (R7081-100), 36 are 13-inch (R8055), and 14 are 20-inch (R7250), all manufactured by Hamamatsu \cite{Hamamatsu}. 
Light pipes are installed between the LS vessel and PMTs to increase the efficiency of photon collection.
All components of the apparatus are mounted inside a cylindrical water tank, with a height of 4~m and a diameter of 3~m.
To reduce the ($n, \gamma$) background in the detector materials and rocks, a passive shield consisting of lead blocks and silicon rubber sheets containing boron carbide (B$_4$C) is installed both inside and outside the tank \cite{Nakajima_shield}. 
More details of the detector setup can be found in \cite{ajimura2020low}.
\par
The data acquisition (DAQ) system consists of 74~channels of 500~MHz-8~bits-8~buffers flash analog-to-digital converters (FADCs), 
of which 62 are used for recording PMT waveforms and 12 are used for trigger purposes \cite{Khai2019}. 
Trigger logics implemented in the global trigger control of our DAQ system include a dual-gate trigger to select the CaF$_2$ signal, and other trigger logics for monitoring purposes (a clock trigger of 3~Hz, a minimum bias trigger to select LS signals, a low-threshold dual-gate trigger to select CaF$_2$ signals at a lower threshold, and a cosmic-ray trigger) \cite{IEEE-Maeda}.
The clock trigger of 3~Hz is used to study single photoelectron charges in dark current and baseline fluctuation, which are discussed in Section \ref{sec:ErrCharge}.
\subsection{$2\nu\beta\beta$ and energy resolution}
\label{ssec:2nbb}
\begin{figure}[h!]
	\centering
	\includegraphics[page=1,width=0.9\linewidth]{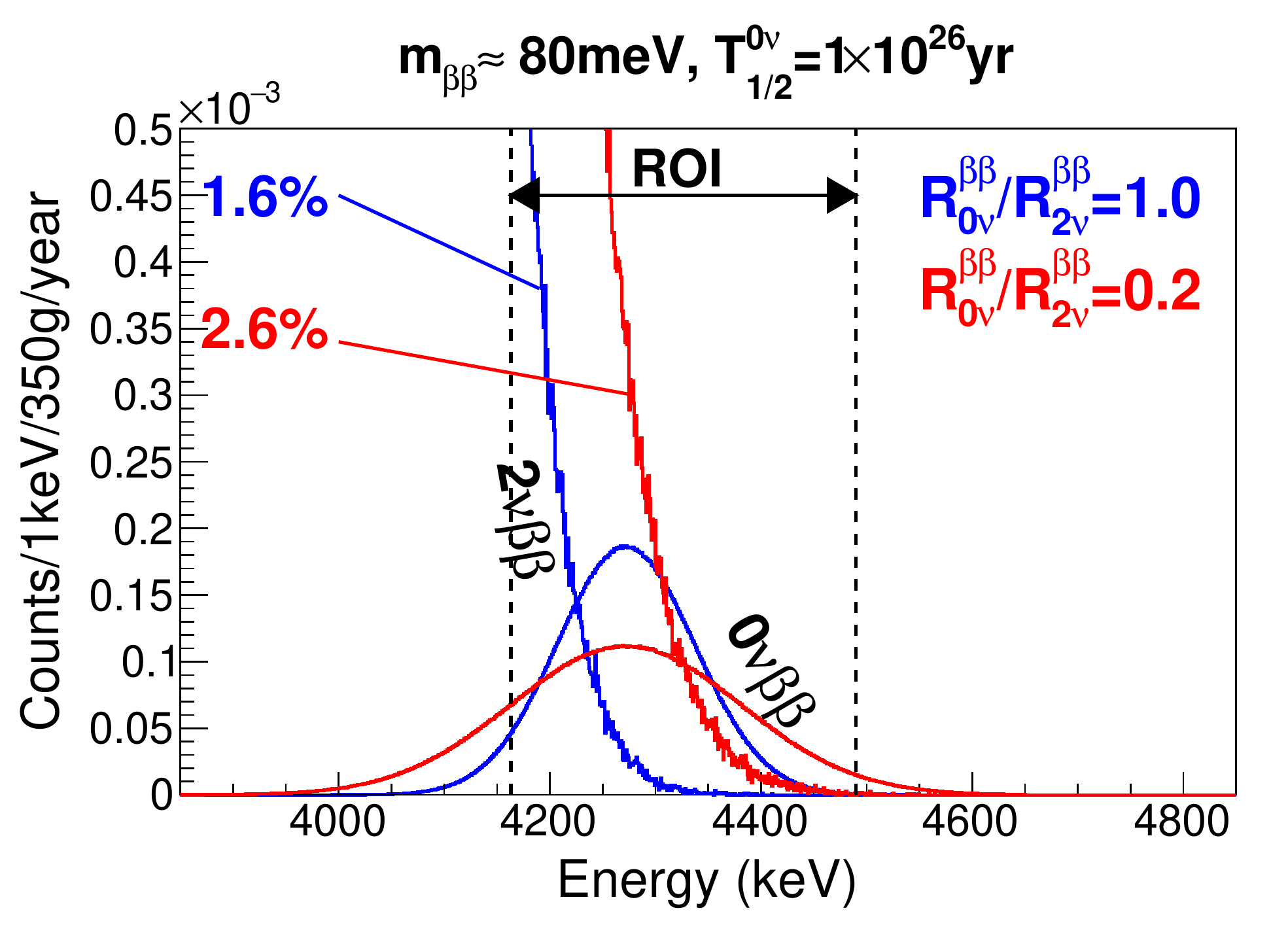}
	\caption{Simulation histograms of $0\nu\beta\beta$ and $2\nu\beta\beta$ of $^{48}$Ca with energy resolutions of 2.6$\%$ (red) and 1.6$\%$ (blue). 
		The ROI to compare the ratio of $0\nu\beta\beta$ to $2\nu\beta\beta$ ($R^{\beta\beta}_{0\nu}$/$R^{\beta\beta}_{2\nu}$) is marked with dashed black lines.
	}
	\label{fig:hist2vbb}
\end{figure}

In $0\nu\beta\beta$ experiments, $2\nu\beta\beta$ is an irremovable background proportional to the mass of $^{48}$Ca.
We plan to develop a ton-scale and highly-enriched-$^{48}$Ca detector to improve the sensitivity of $0\nu\beta\beta$ study, so that $2\nu\beta\beta$ will provide a severe background compared to the $0\nu\beta\beta$.
The energy distributions of $2\nu\beta\beta$ and $0\nu\beta\beta$ of $^{48}$Ca are a continuous spectrum and a mono-energetic peak at $Q_{\beta\beta}$, respectively. 
Figure~\ref{fig:hist2vbb} shows simulation energy spectra of $0\nu\beta\beta$ and $2\nu\beta\beta$ with different energy resolutions.
In this simulation, the $2\nu\beta\beta$ half-life ($T_{1/2}^{2\nu}$) is 4.2$\times$10$^{19}$ years \cite{2vbb} and $0\nu\beta\beta$ half-life ($T_{1/2}^{0\nu}$) is assumed to be 10$^{26}$ years.
This $T_{1/2}^{0\nu}$ is equivalent to an effective neutrino mass (m$_{\beta\beta}$) of 80~meV, which can be obtained by using $T_{1/2}^{0\nu}\propto\left({\rm m}_{\beta\beta}\right)^{-2}$ equation indicated in {\cite{Rev0nbb}}.
This m$_{\beta\beta}$ is close to the world’s-best upper limit of m$_{\beta\beta}$ reported by the \underline{Kam}ioka \underline{L}iquid-scintillator \underline{A}nti-\underline{N}eutrino \underline{D}etector - \underline{Ze}ro \underline{n}eutrino double beta decay search (KamLAND-Zen) collaboration \cite{PhysRevLett.117.082503}.
Improving the resolution from 2.6$\%$ to 1.6$\%$ increases the ratio of $0\nu\beta\beta$ to $2\nu\beta\beta$ from 0.2 to 1.0 within the region of interest (ROI); hence, a better energy resolution is required to reduce the $2\nu\beta\beta$ background.
\par
The CaF$_2$ signal consists of many photoelectrons (p.e.). In an ideal case, the energy resolution should be equal to the statistical fluctuation of the number of p.e. ($N_{\rm p.e.}$). 
At the $Q_{\beta\beta}$ of $^{48}$Ca, the current energy resolution of the CANDLES III detector is 2.6$\%$ \cite{Ohata_DThesis}, and with a p.e. yield of 0.91 p.e./keV, the statistical fluctuation of $N_{\rm p.e.}$ is 1.6$\%$. 
The resolution is larger than the statistical fluctuation, and other fluctuations are likely to be present that degrade the resolution further. 
The fluctuations affecting the CANDLES energy resolution include statistical fluctuation, detector stability, crystal position, and charge measurement.
Statistical fluctuation is mainly influenced by $N_{\rm p.e.}$; hence, we cool the CaF$_2$ crystals at approximately 5~$^{\circ}$C to increase the scintillation light output, install light pipes to increase the photo-coverage \cite{CANDLES_Umehara_EPJ_2014}, and introduce a magnetic cancellation coil to increase the efficiency of p.e. collection \cite{CANDLES_IIDA_Taup}.
The detector stability and crystal position were studied in previous researches \cite{Ohata_DThesis,CANDLES_IIDA_Taup} and were found to slightly influence the energy resolution of CANDLES.
In this paper, the uncertainty in charge measurement of the CANDLES III detector is discussed (Section \ref{sec:ErrCharge}), and a method to reduce the uncertainty is introduced (Section \ref{sec:PC}).

\section{Error in charge measurement}
\label{sec:ErrCharge}
\begin{figure}[b!]
	\centering
	\includegraphics[page=1,width=0.9\linewidth]{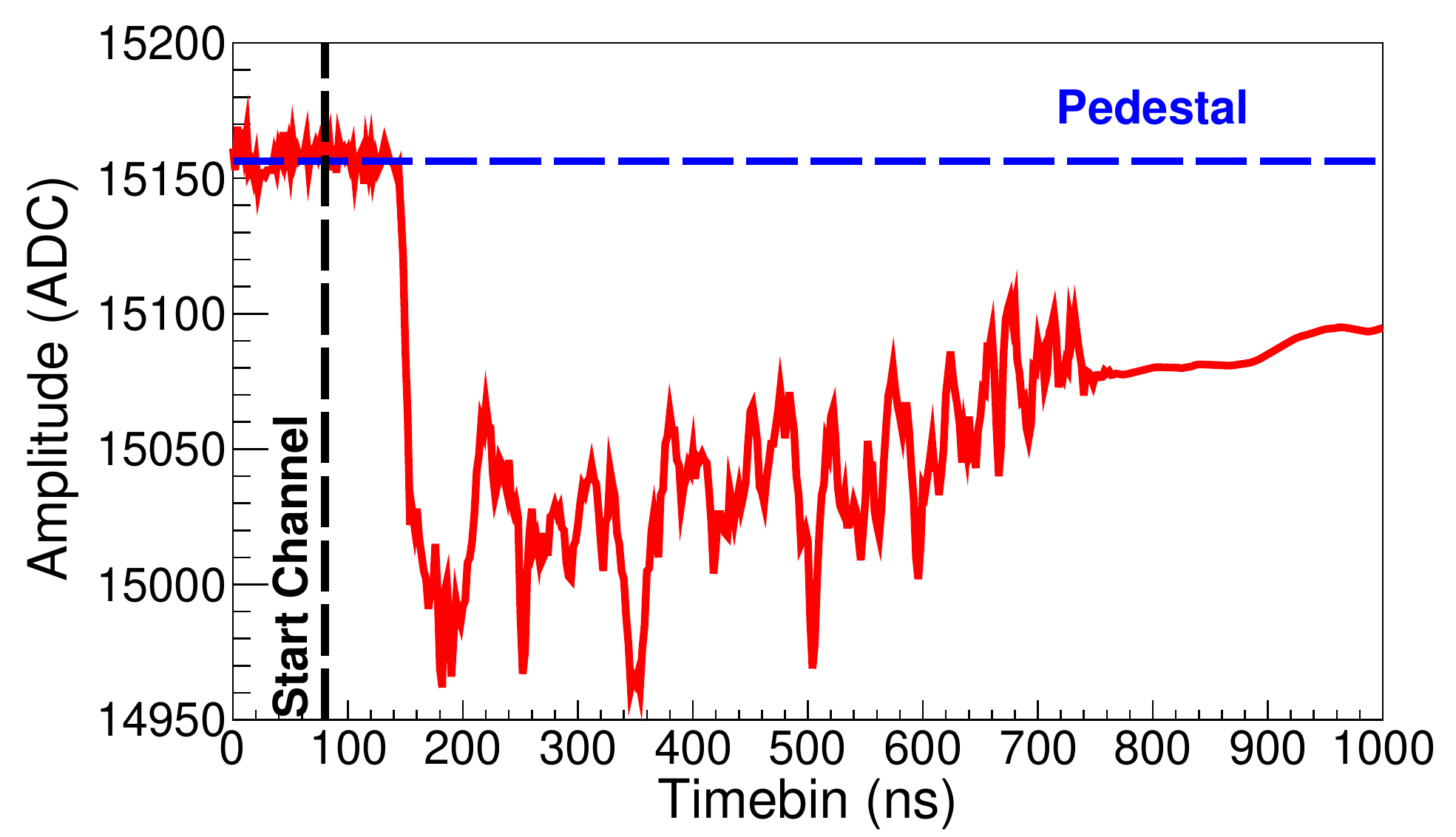}
	\caption{CaF$_2$ sum waveform of 62 PMTs in CANDLES. 
		The values at first bins, before the Start Channel, are used to calculate the pedestal.
	}
	\label{fig:waveform}
\end{figure}
In the current analysis, we sum the waveforms of 62~PMTs and calculate the charge using signal integration = $\rm{\Sigma_{\textit i}}(Pedestal-Signal[{\textit i}])$, where $i$ is the waveform's time bin, with each time bin equivalent to 2~ns.
Figure~{\ref{fig:waveform}} shows the sum waveform of 62 PMTs of a CaF$_2$ signal.
The dashed black line indicates the Start Channel of the waveform.
Digitized values before the Start Channel are used to calculate the pedestal, which is indicated by a dashed blue line.
The discrete and smooth regions are because of fine and sum sample recording, which is discussed in section {\ref{ssec:setup}}.
Because of CaF$_2$'s long decay constant of 1000~ns, each signal is integrated in a time interval of 4000~ns; hence, the baseline fluctuations are accumulated. 
In this study, we examined possible fluctuations over a long interval, including dark current in the PMTs, noise in the baseline, digitization error (related to the resolution of the FADCs), and pedestal uncertainty.
In the following subsections, these long-interval fluctuations at a $\gamma$-peak of $^{40}$K (1460.8~keV), a $\gamma$-peak of $^{208}$Tl (2614.5~MeV), and $Q_{\beta\beta}$ of $^{48}$Ca are estimated.

\subsection{Dark current}
\label{ssec:DC}
\begin{figure}[h!]
	\centering
	\includegraphics[width=0.99\linewidth]{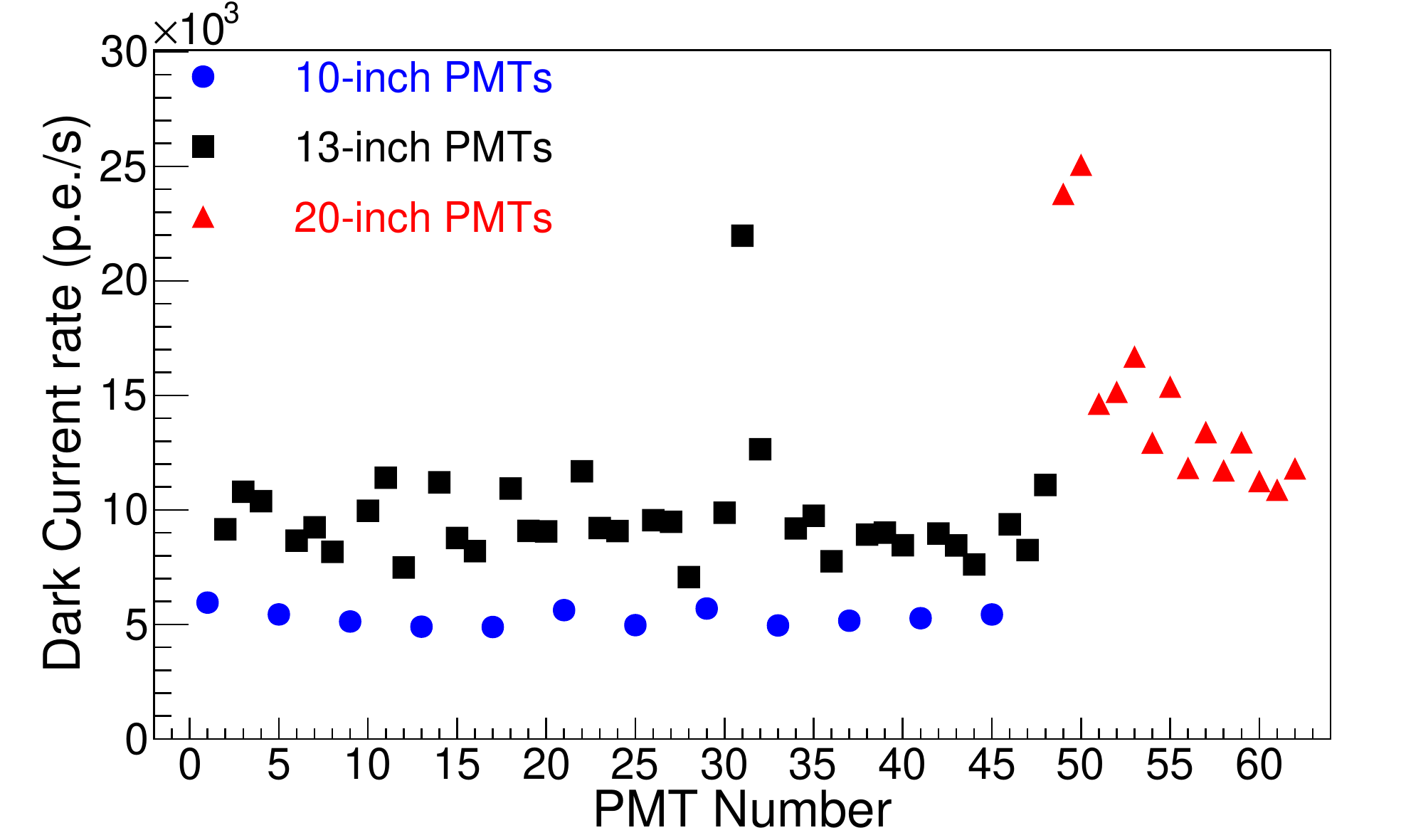}
	\caption{Dark current rates of 10-inch (12, blue circles), 13-inch (36, black squares), and 20-inch PMTs (14, red triangles).}
	\label{fig:DarkRate}
\end{figure}

Dark current (DC) can be contaminated in a CaF$_2$ waveform, affecting the energy resolution statistically. 
For DC analysis, we use the clock trigger events to avoid the p.e. from the scintillator, but the p.e. signals from scintillation photons of low energy radiations may be accidentally collected.
To estimate the DC rate in each PMT, a threshold is individually set for each PMT to count the p.e. signals. 
Details of the threshold setting are described in Subsection~{\ref{ssec:overlap}}.
Figure~\ref{fig:DarkRate} shows the DC rates of 62~PMTs, all of which are in the order of 10$^4$~p.e./s.
From summing the DC rates of 62~PMTs, the DC rate in the sum waveform is approximately $\mu_{\rm{DC}}$=6.2$\times$10$^{5}$~p.e./s. 
Thus, the DC accumulated in an integration interval $T_{\rm INT}$ is calculated as $N_{\rm{DC}}$=$\mu_{\rm{DC}}\times T_{\rm INT}$, and the fluctuation of DC is $\sigma_{\rm{DC}}$=$\sqrt{N_{\rm{DC}}}$.
In 4000~ns of the summed waveform, the average amount of DC is 2.5 p.e., and the fluctuation of DC is $\sigma_{\rm{DC}}$=1.6 p.e. 
The relative uncertainties induced by DC ($\sigma_{\rm DC}$/$N_{\rm p.e.}$) at $^{40}$K, $^{208}$Tl, and $Q_{\beta\beta}$($^{48}$Ca) are 0.1$\%$, 0.06$\%$, and 0.04$\%$, respectively.
The $N_{\rm p.e.}$ at the $^{40}$K peak, $^{208}$Tl peak, and $Q_{\beta\beta}$($^{48}$Ca) can be calculated from the p.e. yield.

\subsection{Noise in baseline}
\label{ssec:noise}
\begin{figure}[h!]
	\centering
	\includegraphics[width=\linewidth,page=1,clip,trim={0mm 0mm 0mm 0mm}]{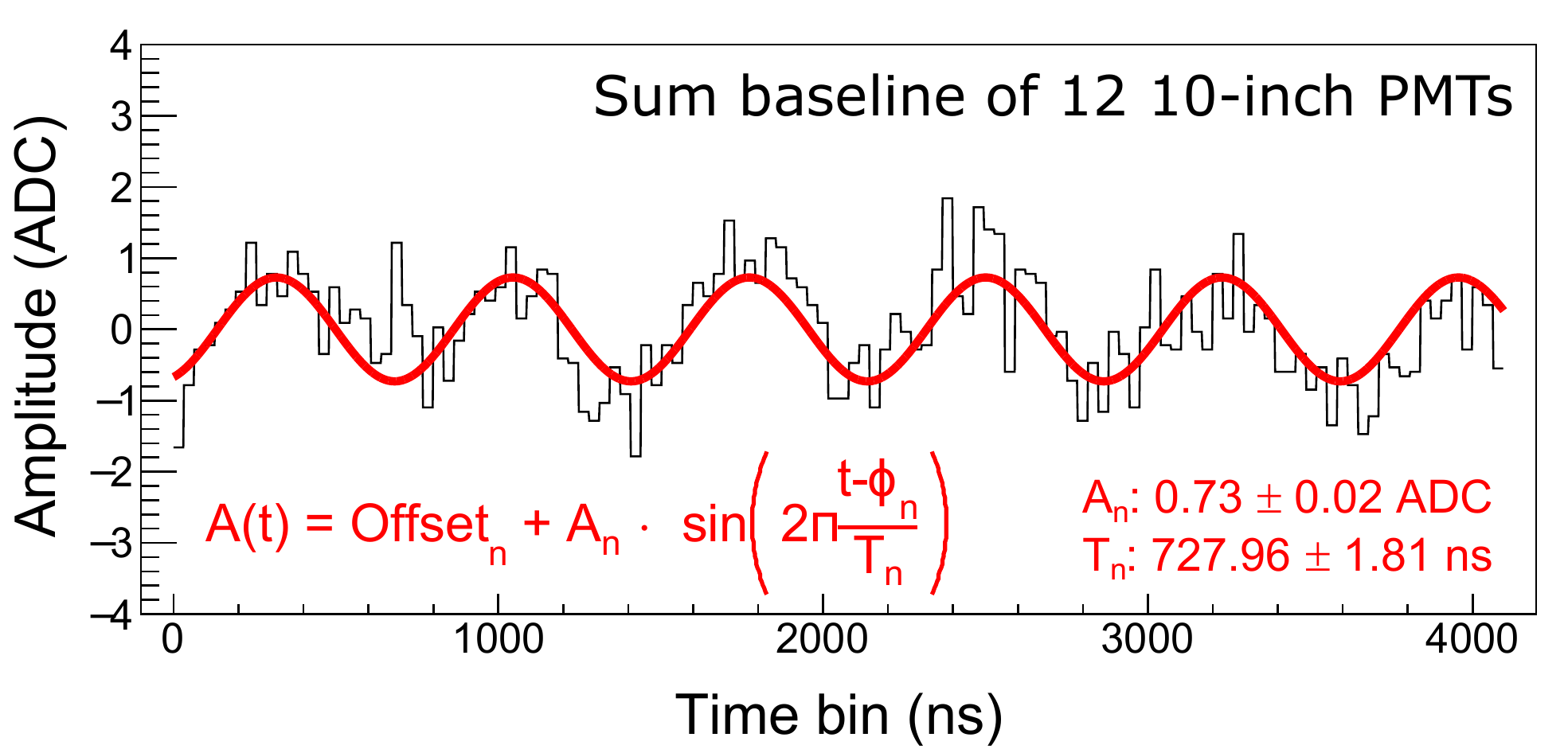}
	\caption{Sine fitting on the sum baseline of 12 10-inch PMTs in one clock event of CANDLES III.}
	\label{fig:pmtNoise}
\end{figure}

The sine-wave noise of 62~PMTs is analyzed using a clock trigger, with visible noise found in the baseline of the 10-inch PMTs, whereas the noise amplitudes of the 13-inch and 20-inch PMTs are not visible.
We sum the 12 baselines of 10-inch PMTs for noise analysis and estimate the effect of noise on the energy resolution.
Figure~\ref{fig:pmtNoise} shows a sinusoidal shape of the sum baseline of the 12 10-inch PMTs in one clock event.
For every clock event, the sum baseline is fitted with a sine function to estimate the noise amplitude and cycle:
\begin{equation}
	A(t) = A_{\rm n} {\rm sin}\left(2\pi\frac{t-\varphi_{\rm n}}{T_{\rm n}}\right),
\end{equation}
where $A_{\rm n}$ is the noise amplitude, $T_{\rm n}$ is the noise cycle, and $\varphi_{\rm n}$ is the phase factor.
From analyzing more than 10$^5$ clock events, the mean values of the noise amplitude and cycle are found to be 0.73~ADC, equivalent to approximately 3~mV in the FADC input, and 730~ns, respectively.
The noise amplitude is small, although we sum the noise of the 12 10-inch PMTs. 
Each PMT signal is amplified by 10 times before being fed into an FADC, implying that the noise amplitude is approximately 0.3~mV before amplification. 
The impedance of each PMT is 50~$\Omega$; therefore, the power of the sine-wave noise is only 1.8~nW, which is extremely small.
This noise is caused by the noise resonance that varies with frequency, depending on the length of the coaxial cable.
The 10-inch PMTs' cable length, which differs from those of 13-inch and 20-inch PMTs, is one quarter of the 730~ns noise's wavelength, 
and therefore, the 10-inch PMT's cable supports this particular noise resonance. 
However, as its effect is negligible, the source of this sinusoidal noise has not been studied in this paper.
The fluctuation induced by sinusoidal noise in the signal integration in $T_{\rm INT}$~ns is as follows:
\begin{equation}
	\begin{split}
		\rm{\sigma_{noise}}&=\int_{0}^{{T}_{\rm INT}}A(t)dt\\
		&=\frac{A_{\rm n}T_{\rm n}}{\pi\bar{\mu}_{\rm p.e.}}{\rm sin}\left(\pi\frac{T_{\rm{INT}}}{T_{\rm n}}\right){\rm sin}\left(\pi\frac{T_{\rm{INT}}-2\varphi_{\rm n}}{T_{\rm n}}\right),
	\end{split}
	\label{eq:sigmaNoise}
\end{equation}
where $\rm{\bar{\mu}_{p.e.}}$, which is the average charge of single-p.e. (1~p.e.) signals of the 62~PMTs in the ADC unit, is used to convert $\rm{\sigma_{noise}}$ into p.e. units.
The noise effect is clearly a sine function of phase $\rm{\varphi_{n}}$, which is random in every CaF$_2$ event and difficult to estimate.
Thus, we estimate the maximum fluctuation induced by the 730-ns-cycle sine noise in this research.
The maximum fluctuation, which is the root mean square of the sine function in equation \ref{eq:sigmaNoise}, is a function of $T_{\rm INT}$:
\begin{equation}
	\sigma^{\rm max}_{\rm noise}(T_{\rm INT}) = \frac{A_{\rm n}T_{\rm n}}{\sqrt{2}\pi\bar{\mu}_{\rm p.e.}}\left|{\rm sin}\left(\pi\frac{T_{\rm INT}}{T_{\rm n}}\right)\right|.
	\label{eq:sigmaHF}
\end{equation}
For the integration interval of 4000~ns, the maximum effect of the 730-ns-cycle noise is 2~p.e.; hence, the maximum relative uncertainties of noise ($\sigma^{\rm{max}}_{\rm{noise}}/{N_{\rm p.e.}}$) at $^{40}$K, $^{208}$Tl, and $Q_{\beta\beta}$($^{48}$Ca) are 0.15$\%$, 0.08$\%$, and 0.05$\%$, respectively.

\subsection{Digitization error}
\label{ssec:DigitErr}
\begin{figure}[b!]
	\centering
	\includegraphics[page=1, width=0.9\linewidth]{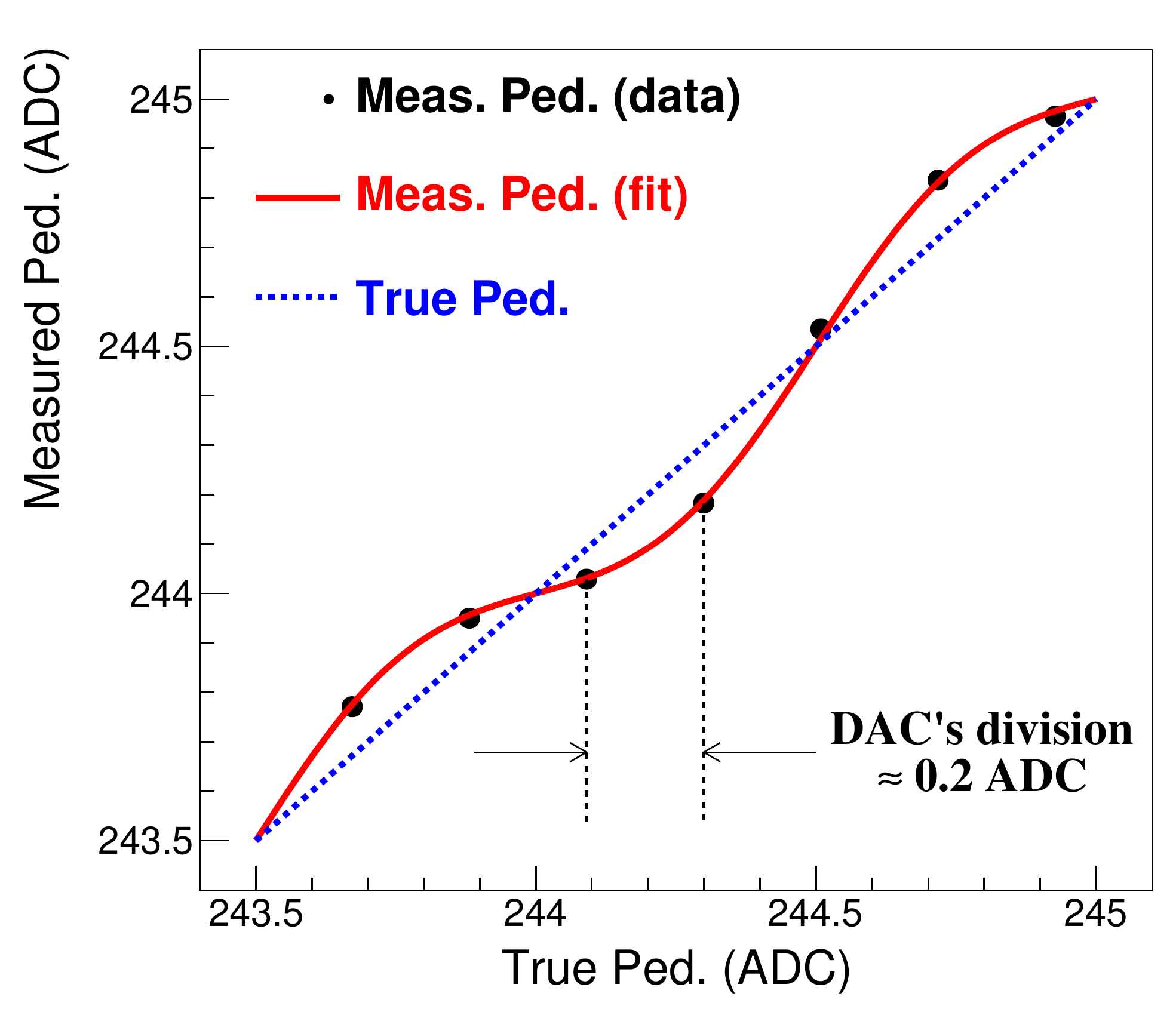}
	\caption{
		Measured pedestal plotted as a function of true pedestal with three $\sigma_{\rm{A}}$ values.
	}
	\label{fig:DigitErr-1}
\end{figure}

We record the PMT waveform using a 500~MHz-8~bits ADC08DL502 from Texas Instruments with 7.5 effective number of bits \cite{TI_ADC}.
The probability of recording a digitized value ``$\rm n$" is given as the integral within the range $\rm n \pm 0.5$ of a Gaussian function $P_{\rm n}(\mu_{\rm A},\sigma_{\rm A})$, where $\mu_{\rm A}$ and $\sigma_{\rm A}$ are the mean and standard deviation values, respectively, of the analog input. 
In this study, the pedestal is calculated as the average value of the first 40~data~points, equivalent to 80~ns, in the waveform.
The measured pedestal is theoretically calculated as $\Sigma_{\rm n}({\rm n}\times P_{\rm n})$.
Figure~\ref{fig:DigitErr-1} shows the measured pedestal as a function of true pedestal, or $\mu_{\rm A}$, in one PMT.
To perform the test, we adjust the true pedestal using a 12-bit digital-to-analog converter (DAC), which is installed in each FADC module, with its voltage division equivalent to approximately 0.2~ADC units. 
The black points depict experimental data obtained at different true pedestal values, the solid red line is the fitting function, and the blue dashed line is the expected true pedestal.
Because of the least significant bit (LSB), the measured pedestal is different from the true pedestal, and this difference is called the digitization error (DE). 
The DE causes a fluctuation in the measured pedestal, which accumulates when calculating the signal integration.
In a 1~p.e. signal, the DE accumulates at the non-pedestal points in the width of the 1~p.e. signal.
According to Figure~{\ref{fig:DigitErr-1}}, the DE is zero at the pedestal of 244.5~ADC; thus, the 1~p.e. charge at this pedestal value is not affected by the DE. 
Because the measured pedestal of each PMT is set at nearly 244.5~ADC, the DE fluctuation on the 1~p.e. charge of one PMT can be assumed to be linearly correlated with the measured pedestal:
\begin{equation}
	\Delta^{\rm iPMT}_{\rm 1 p.e.} = Slope^{\rm iPMT}\times \Delta^{\rm iPMT}_{\rm Ped.},
	\label{delta1pe}
\end{equation}
where $Slope^{\rm iPMT}$ is the linear coefficient and $\rm{\Delta^{iPMT}_{Ped.}}$ is equal to $\rm{Pedestal^{iPMT}-244.5\;ADC}$.
The DE is accumulated at non-pedestal points; if the 1~p.e. signals do not overlap in high-$N_{\rm p.e.}$ events, the number of non-pedestal points is $N_{\rm p.e.}w/\delta$, where $w$ is the width of the 1~p.e. signal, and $\delta$ is the FADC sampling interval (2~ns).
In contrast, if many 1~p.e. signals overlap each other, the number of non-pedestal points is reduced.
In this study, we estimate the reduction factor using the following mathematical model. 
The CaF$_2$ waveform follows an exponential function:
\begin{equation}
	\mu(t)=\mu(0)e^{-t/\tau}=\frac{N_{\rm p.e.}}{\tau}e^{-t/\tau},
	\label{eq:waveform}
\end{equation}
where $\mu(t)$ is the signal amplitude at time $t$, and $\tau$ is the decay constant of CaF$_2$, which is 1000~ns.
The expected number of p.e. within the width of the 1~p.e. signal is $\mu(t)w$, and the probability of obtaining the pedestal point is $q(t)={\rm e}^{-\mu(t)w}$.
The number of non-pedestal points can be deduced as follows:
\begin{equation}
	N_{\rm signal}= \frac{1}{\delta}\int_{0}^{T_{\rm INT}}(1-q(t))dt,
\end{equation}
and the reduced factor is estimated as
\begin{equation}
	R=\frac{N_{\rm signal}}{N_{\rm p.e.}w/\delta}.
\end{equation}
\par
\begin{figure}[b!]
	\centering
	\includegraphics[page=1, width=0.85\linewidth, clip, trim={0mm 0mm 0mm 0mm}]{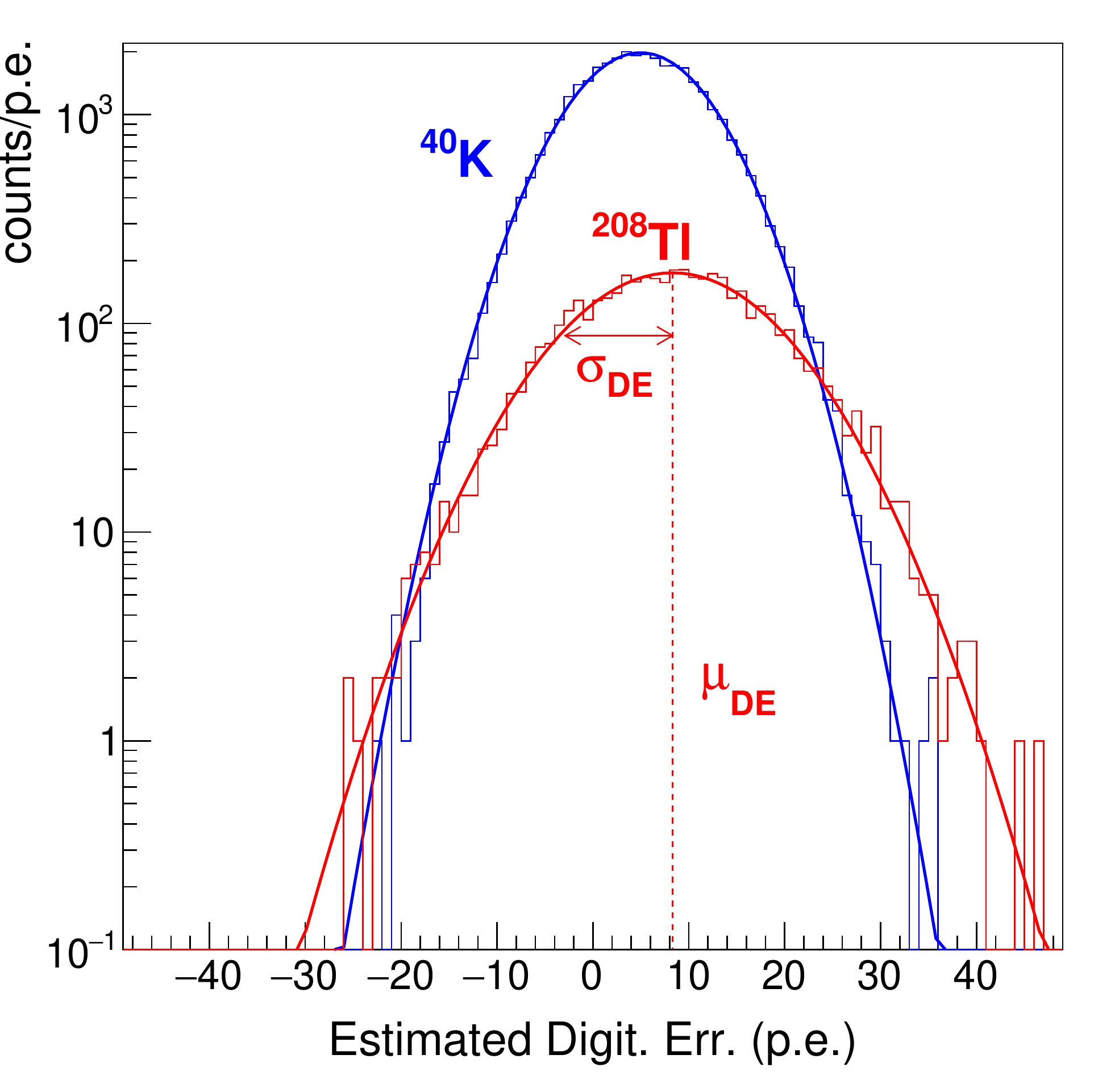}
	\caption{
		Distributions of estimated DEs in 62~PMTs with Gaussian fitting at the $\gamma$-peaks of $^{40}$K (blue) and $^{208}$Tl (red).
	}
	\label{fig:DigitErr-2}
\end{figure}

Each CaF$_2$ waveform contains up to several thousands of 1~p.e. signals from the 62~PMTs.
Because the pedestal and number of p.e. are not the same in every PMT, the DE is estimated individually for each PMT using the corresponding $n^{\rm iPMT}_{\rm p.e.}$ and measured pedestal in that PMT:
\begin{equation}
	\omega^{\rm iPMT} = n^{\rm iPMT}_{\rm p.e.}\times\Delta^{\rm iPMT}_{\rm 1 p.e.}\times R^{\rm iPMT}, 
	\label{DEpmt}
\end{equation}
and the DE of the whole detector is the sum of DEs in the 62~PMTs: $\rm{\Omega}=\rm{\sum_{iPMT=1}^{62}{\omega^{iPMT}}}$.
In Figure~\ref{fig:DigitErr-2}, the distributions of estimated summed DEs of 62~PMTs on the $\gamma$-peaks of $^{40}$K and $^{208}$Tl with the Gaussian fitting functions are plotted in blue and red, respectively.
For the DE distribution at each energy peak, the mean value of DE ($\mu_{\rm{DE}}$) causes a shift in the mean energy peak, and the standard deviation ($\sigma_{\rm{DE}}$) influences the energy resolution.
As shown in the above equations, the DE value corresponds to the measured pedestal and $N_{\rm p.e.}$, which depends on the energy and integration interval.
The standard deviation of each distribution is due to the fluctuation of measured pedestal and statistical fluctuation of $N_{\rm p.e.}$.
The differences in $\mu_{\rm{DE}}$ and $\sigma_{\rm{DE}}$ at $^{40}$K and $^{208}$Tl energy peaks in Figure~{\ref{fig:DigitErr-2}} are because of the difference in $N_{\rm p.e.}$.
With an integration interval of 4000~ns, the $\sigma_{\rm{DE}}$ at the $\gamma$-peaks of $^{40}$K and $^{208}$Tl are 7.3 p.e. and 10.4 p.e., respectively, and the relative fluctuation of DE ($\sigma_{\rm{DE}}/{N_{\rm p.e.}}$) on these $\gamma$-peaks are 0.55$\%$ and 0.44$\%$, respectively.
According to these obtained fluctuations, the relative fluctuation induced by DE on the $Q_{\beta\beta}$($^{48}$Ca) should be small.

\subsection{Pedestal uncertainty}
\label{ssec:PedErr}
\begin{figure}[h!]
	\centering
	\includegraphics[width=0.9\linewidth, page=1,clip,trim={0mm 0mm 0mm 0mm}]{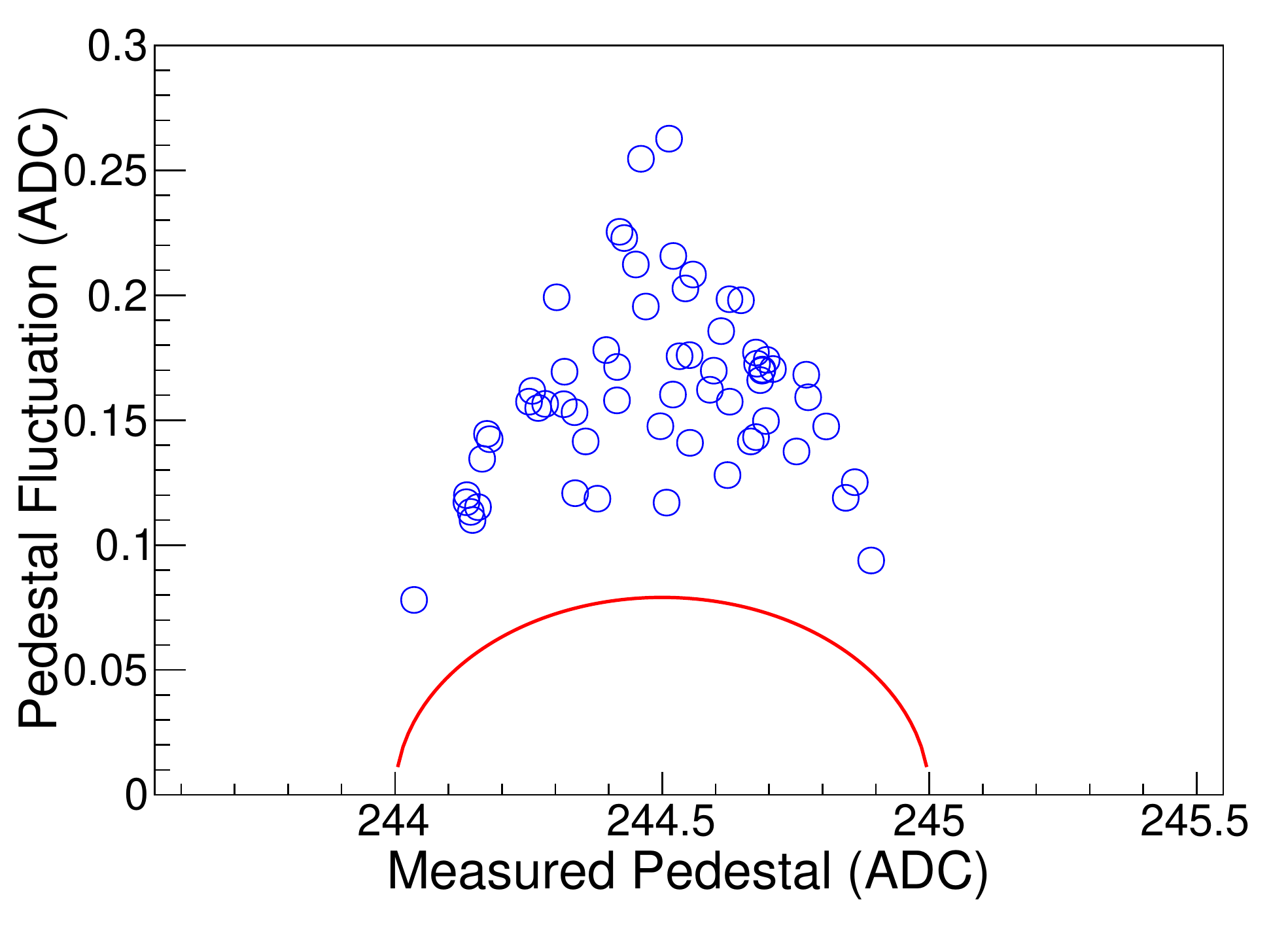}
	\caption{
		Standard deviations of measured pedestal distributions plotted with blue points as a function of measured pedestal of the 62~PMTs.
		The solid red line is the expected binomial fluctuation.}
	\label{fig:SigPedBino}
\end{figure}

In this study, the pedestal of each PMT is calculated by taking the average of the first 40 sampling points in the waveform. 
Because the pedestal of each PMT is adjusted to approximately 244.5~ADC, it follows a binomial distribution of 40 trials with the obtained value in each trial as either 244 or 245~ADC. 
Taking $p$ as the probability of obtaining 245~ADC in a single trial in one PMT, the statistical uncertainty of the pedestal in that PMT is
\begin{equation}
	\sigma^{\rm iPMT}_{\rm PedStat}=\sqrt{p(1-p)/N},
	\label{PedFlucStat}
\end{equation}
where $N$ is the number of sampling points used to calculate the pedestal, and $N$ is equal to 40 in the current analysis.
$\sigma^{\rm iPMT}_{\rm PedStat}$ plotted as a solid red line in Figure~\ref{fig:SigPedBino} is the ideal pedestal uncertainty, and the experimental pedestal uncertainty ($\sigma^{\rm iPMT}_{\rm Ped}$) in each PMT is obtained by taking the root mean square of the measured pedestal distribution of each PMT. 
The blue circles in Figure~\ref{fig:SigPedBino} are the experimental pedestal fluctuations of 62~PMTs. 
Because our PMTs are affected by noise, $\sigma^{\rm iPMT}_{\rm Ped}$ is approximately twice as large as $\sigma^{\rm iPMT}_{\rm PedStat}$.
\begin{figure}[h!]
	\centering
	\includegraphics[width=0.9\linewidth, page=1,clip,trim={0mm 0mm 0mm 0mm}]{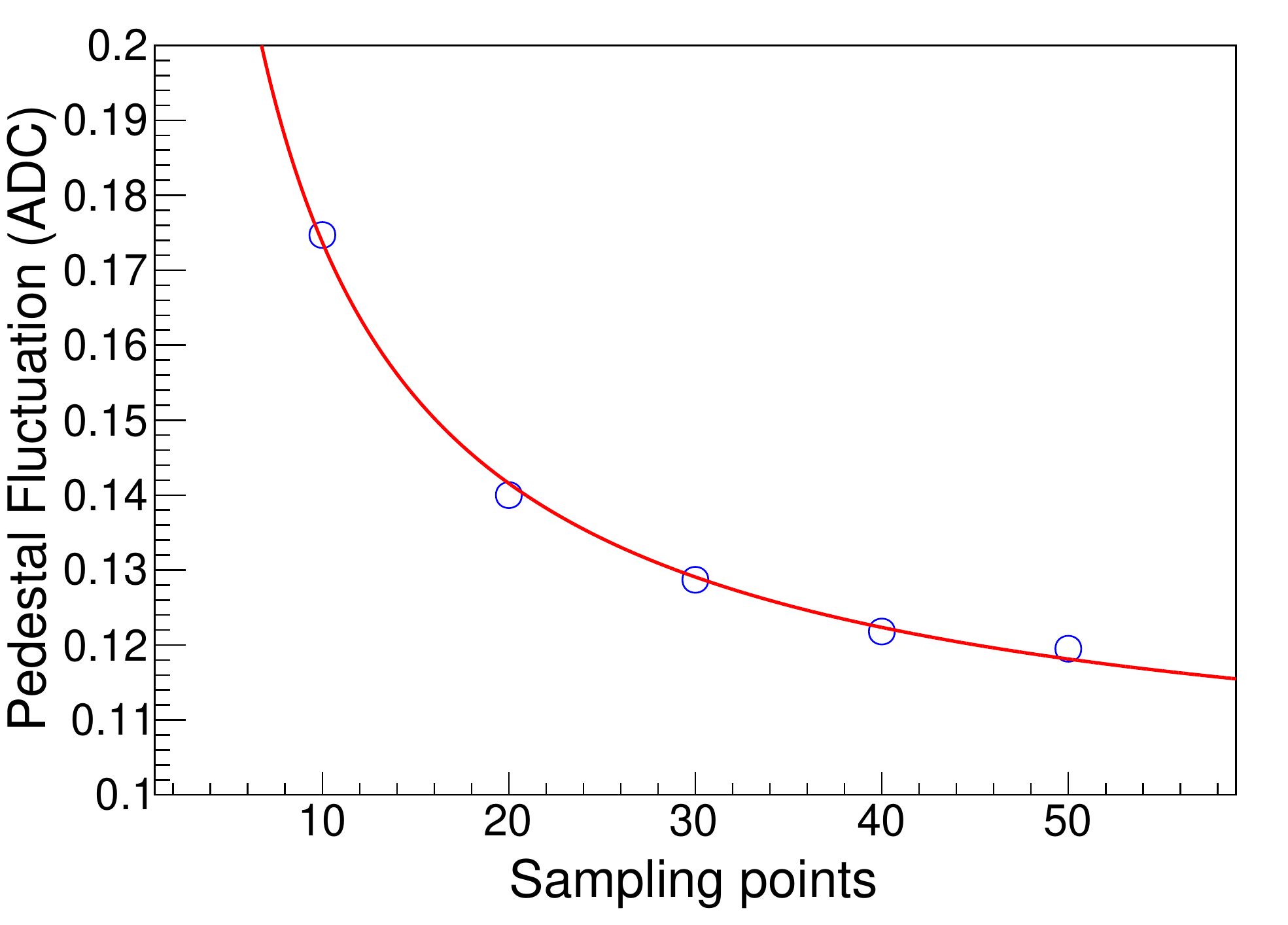}
	\caption{Pedestal fluctuation of one random PMT plotted as a function of number of sampling points. 
		The solid red line is the fitting function $\sqrt{a/N+b}$.
	}
	\label{fig:PedFlucPMT04}
\end{figure}
In Figure~{\ref{fig:PedFlucPMT04}}, $\sigma_{\rm Ped}$ of one random PMT is obtained with a different number of sampling points plotted as blue circles.
Assuming that $\sigma_{\rm Ped}$ is a root sum square of $\sigma_{\rm PedStat}$ and noise fluctuation, a function $\sqrt{\sigma^{2}_{\rm PedStat}+\sigma^{2}_{\rm noise}}=\sqrt{a/N+b}$ used for fitting is plotted as a solid red line in Figure~{\ref{fig:PedFlucPMT04}}.
With more number of sampling points, $\sigma_{\rm Ped}$ is reduced because of the improvement of $\sigma_{\rm PedStat}$.
The saturation of this $\sigma_{\rm Ped}$ from 40~points may be caused by the increased noise fluctuation.
\par
\begin{figure}[h]
	\centering
	\includegraphics[width=0.9\linewidth, page=1,clip,trim={0mm 0mm 0mm 0mm}]{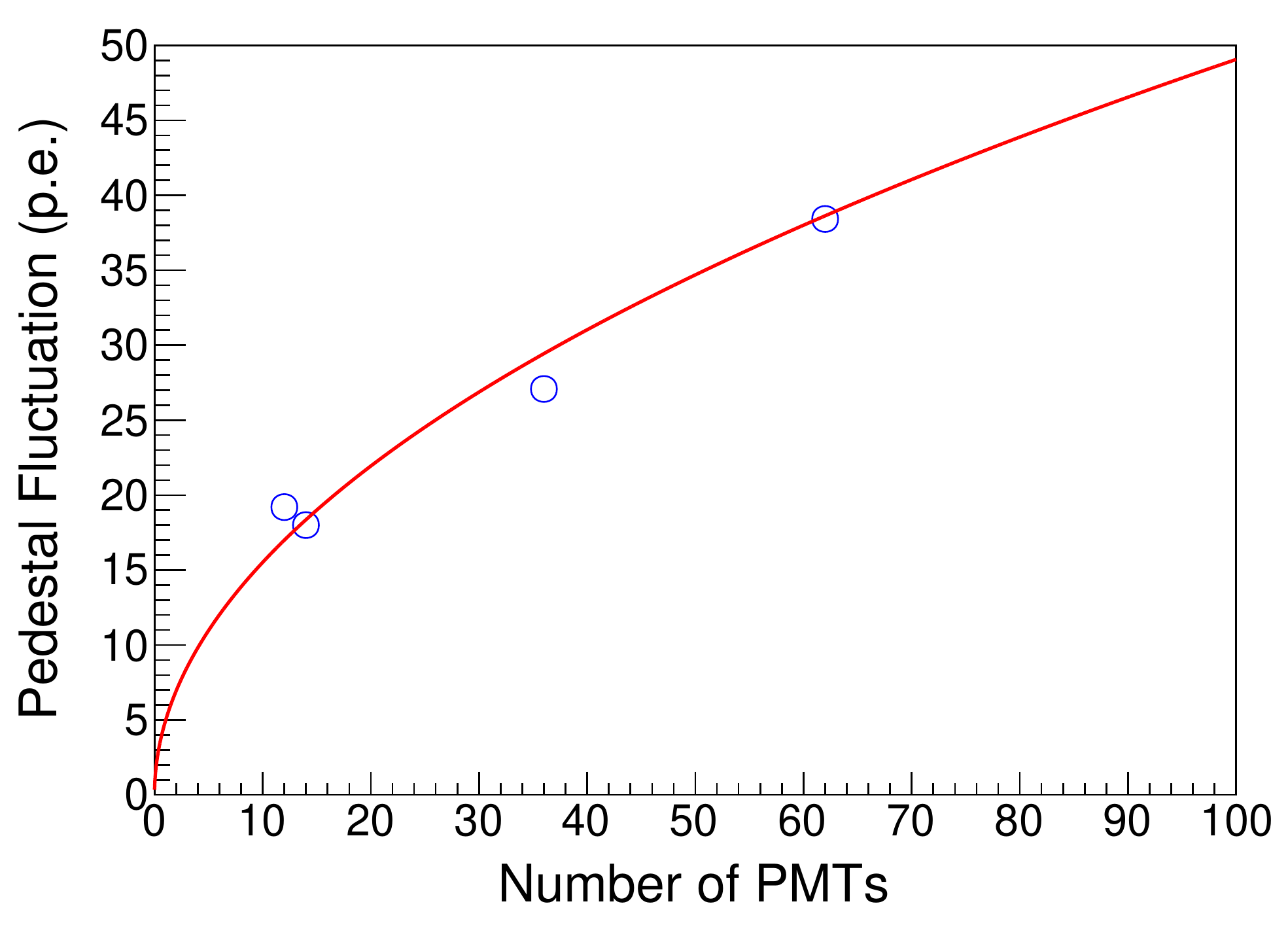}
	\caption{Pedestal fluctuation as a function of number of PMTs in different integration intervals.
		From left to right, respectively, the summed pedestal fluctuations of twelve~10-inch~PMTs, fourteen~20-inch~PMTs, thirty-six~13-inch~PMTs and all 62~PMTs are plotted as blue circles.
		The solid red line is an estimated function $\sqrt{{N_{\rm PMT}}}\times\bar{\sigma}_{\rm{PedErr}}$, where $N_{\rm PMT}$ is the number of PMTs and $\bar{\sigma}_{\rm{PedErr}}$ is the average pedestal fluctuation of 62 PMTs.
	}
	\label{fig:PedFlucPMT}
\end{figure}

In signal integration, the pedestal uncertainty is accumulated at every data point, and the accumulated fluctuations ($\sigma^{\rm iPMT}_{\rm PedErr}$) are linearly proportional to the number of sampling points ($N'$): $\sigma^{\rm iPMT}_{\rm PedErr}=N'\times\sigma^{\rm iPMT}_{\rm Ped}$.
The pedestals of the 62~PMTs are summed in the sum waveform, and the fluctuation induced by the pedestal uncertainty of the 62~PMTs is amplified by several times: 
\begin{equation}
	\sigma_{\rm PedErr} = \sqrt{\sum_{\rm  iPMT=1}^{62}{\left(\sigma^{\rm iPMT}_{\rm PedErr}\right)^2}},
\end{equation}
and $\sigma_{\rm PedErr}$ is also linearly proportional to the number of sampling points. 
Figure {\ref{fig:PedFlucPMT}} shows the summed fluctuation in p.e. unit of twelve~10-inch~PMTs, fourteen~20-inch~PMTs, thirty-six~13-inch~PMTs and all 62~PMTs, which are plotted in blue circles.
The solid red line is an estimated function $\sqrt{{N_{\rm PMT}}}\times\bar{\sigma}_{\rm{PedErr}}$, where $N_{\rm PMT}$ is the number of PMTs, and $\bar{\sigma}_{\rm{PedErr}}=\sigma_{\rm{PedErr}}/\sqrt{62}$ is the average pedestal fluctuation of 62~PMTs.
Because of noise fluctuations, the pedestal fluctuations of 10-inch, 13-inch and 20-inch~PMTs are slightly different from the estimated values.
\par
Replacing the current 8-bit ADCs with 12-bit ADCs can degrade the DE fluctuation ($\sigma_{\rm DE}$) and the pedestal's statistical fluctuation ($\sigma_{\rm PedStat}$) to a negligible level.
The pedestal thus can be measured more precisely, and the effect of the baseline fluctuation is degraded.
As the ADC replacement requires an effort in the development of a DAQ system, we want to minimize the baseline fluctuation in the current setup using waveform analysis, which is discussed in Section {\ref{sec:PC}}.
\par
In this analysis, we make distributions of integration of the baseline for each PMT, and the root mean square (RMS) of the distribution is $\sigma^{\rm iPMT}_{\rm PedErr}$. 
For $\sigma_{\rm PedErr}$, we make the distribution of integration of the sum baseline of the 62~PMTs, and obtain the RMS value. 
$\rm{\sigma_{PedErr}}$ with the current signal integration calculation can be estimated using an integration interval of 4000~ns (or 2000~sampling~points).
In the current analysis, $\rm{\sigma_{PedErr}}$ in each 4000~ns integration of the CaF$_2$ waveform is 38.6~p.e.
The relative fluctuations induced by the pedestal uncertainty of the 62~PMTs, or $\sigma_{\rm PedErr}/N_{\rm p.e.}$, at the $^{40}$K peak, $^{208}$Tl peak, and $Q_{\beta\beta}$($^{48}$Ca) are 2.8$\%$, 1.6$\%$, and 1$\%$, respectively.

\subsection{Summary of baseline fluctuations}
\label{ssec:ErrSummary}
\begin{figure}[h!]
	\centering
	\includegraphics[width=0.99\linewidth, page=1, clip, trim={0mm 0mm 0mm 0mm}]{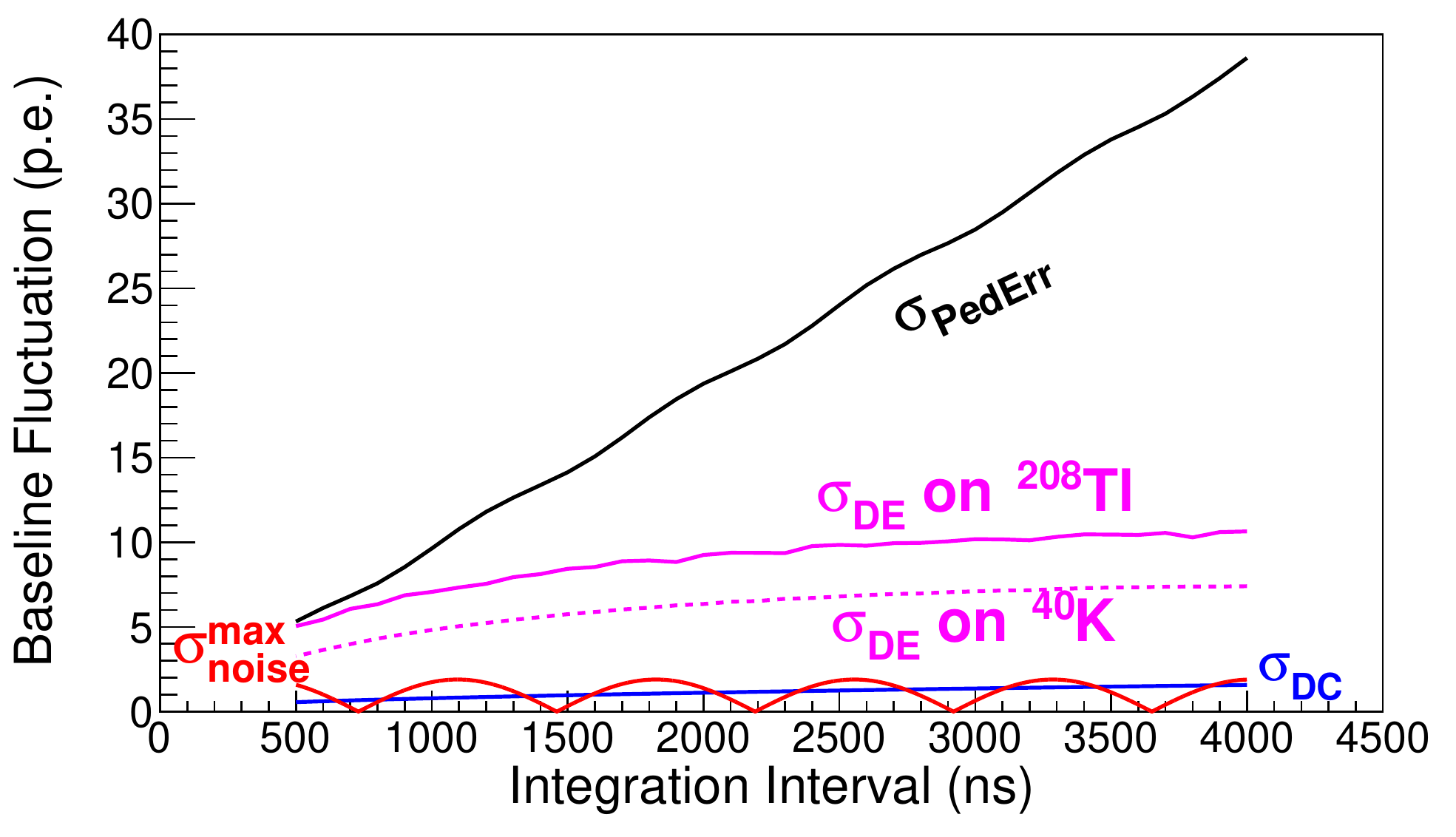}
	\caption{Baseline fluctuations as a function of integration interval.
		Fluctuations induced by DC (blue), maximum fluctuation of 730-ns-cycle noise (red), DE at $\gamma$-peaks of $^{40}$K (magenta dashed) and $^{208}$Tl (solid magenta), and error in pedestal measurement (black).
	}
	\label{fig:BaseFlucPeak}
\end{figure}

The baseline fluctuations are accumulated in the long integration interval of 4000~ns for the CaF$_2$ signal and reduce the energy resolution. 
In this report, we study the baseline fluctuations including DC, sinusoidal noise, DE, and pedestal uncertainty.
These fluctuations are plotted as functions of the integration interval, 500--4000~ns, in Figure~\ref{fig:BaseFlucPeak}.
The DC fluctuation ($\sigma_{\rm{DC}}$), which is proportional to $\sqrt{T_{\rm INT}}$, is plotted with a solid blue line.
The maximum noise fluctuation ($\sigma^{\rm max}_{\rm noise}$), which is calculated using equation \ref{eq:sigmaHF}, is plotted with a solid red line.
The estimated DE fluctuations ($\sigma_{\rm{DE}}$), as functions of the integration interval, at the $\gamma$-peaks of $^{40}$K and $^{208}$Tl are plotted with dashed and solid magenta lines, respectively. 
$\sigma_{\rm DE}$ is proportional to $N_{\rm p.e.}$; therefore, when the integration interval is shrunk, $N_{\rm p.e.}$ is reduced, leading to a reduction in $\sigma_{\rm DE}$.
Because $N_{\rm p.e.}$($^{40}$K) is smaller than $N_{\rm p.e.}$($^{208}$Tl), $\sigma_{\rm DE}$($^{40}$K) is smaller than $\sigma_{\rm DE}$($^{208}$Tl).
By selecting the corresponding energy range in the analysis, $\sigma_{\rm DE}$ at each energy peak is estimated.
The energy resolution is essential in energy range selection, and the energy histograms constructed using partial photon counting with improved resolution (mentioned in {\ref{ssec:PPC}}) are employed. 
The energy histograms constructed with an integration interval less than 500~ns have a poor energy resolution; therefore, the baseline fluctuations are investigated within the 500--4000~ns integration interval.
The fluctuation induced by pedestal uncertainty ($\sigma_{\rm PedErr}$) in the sum waveform of the 62~PMTs is calculated with different integration intervals and plotted as a function of integration interval with a solid black line.
\begin{table}[h!]
	\centering
	\caption{All baseline fluctuations ($\sigma_{\rm DC}$, $\sigma_{\rm noise}$, $\sigma_{\rm DE}$, and $\sigma_{\rm PedErr}$) and statistical fluctuations ($\sigma_{\rm stat}$) at the $^{40}$K peak, $^{208}$Tl peak, and $Q_{\beta\beta}$($^{48}$Ca) with an integration interval of 4000~ns.}
	\begin{tabular}{c c c c}
		& \textbf{$^{40}$K $\gamma$-peak}& \textbf{$^{208}$Tl $\gamma$-peak}& \textbf{$Q_{\beta\beta}$($^{48}$Ca})	\\
		& \textbf{1460.8~keV}	& \textbf{2614.5~keV} 		&  \textbf{4272~keV}			\\
		\hline
		$\sigma_{\rm \textbf{DC}}$									& 1.6 p.e.			& 1.6 p.e. 			&  1.6 p.e.			\\
		($\sigma_{\rm \textbf{DC}}$/$N_{\rm{\textbf{p.e.}}}$)		& (0.1$\%$)			& (0.06$\%$)		&  (0.04$\%$)		\\
		\hline	
		$\sigma_{\rm \textbf{noise}}$								& $\le$2 p.e.		& $\le$2 p.e.		& $\le$2 p.e.		\\
		($\sigma_{\rm \textbf{noise}}$/$N_{\rm \textbf{p.e.}}$)		& ($\le$0.15$\%$) 	& ($\le$0.08$\%$)	& ($\le$0.05$\%$)	\\		
		\hline
		$\sigma_{\rm \textbf{DE}}$									& 7.3 p.e. 			& 10.41 p.e.		&  					\\
		($\sigma_{\rm \textbf{DE}}$/$N_{\rm \textbf{p.e.}}$)		& (0.6$\%$) 		& (0.4$\%$)			& (small)			\\
		\hline
		$\sigma_{\rm \textbf{PedErr}}$								& 38.6 p.e.			& 38.6 p.e. 		& 38.6 p.e.			\\
		($\sigma_{\rm \textbf{PedErr}}$/$N_{\rm \textbf{p.e.}}$)	& (2.9$\%$)			& (1.6$\%$)			& (1.0$\%$)			\\
		\hline
		$\sigma_{\rm \textbf{stat}}=\sqrt{N_{\textbf{\rm p.e.}}}$	& 36.5 p.e.			& 48.8 p.e.			& 62.3 p.e. 		\\
		($\sigma_{{\rm\textbf{stat}}}$/$N_{{\rm\textbf{p.e.}}}$)	& (2.7$\%$) 		& (2.0$\%$)			& (1.6$\%$)			\\		
	\end{tabular}
	\label{tab:BaseFlucContribution}
\end{table}
\par
$\sigma_{\rm PedErr}$ is the most severe, $\sigma_{\rm DE}$ is small, and the fluctuations induced by DC and 730-ns-cycle sinusoidal noise are both negligible.
The baseline and statistical fluctuations ($\sigma_{\rm stat}$) at the $^{40}$K peak, $^{208}$Tl peak, and $Q_{\beta\beta}$($^{48}$Ca) with an integration interval of 4000~ns are listed in Table \ref{tab:BaseFlucContribution}. 
The DE at $Q_{\beta\beta}$ is not estimated, but the relative DE fluctuation should be small. 
$\sigma_{\rm PedErr}$ is a severe fluctuation compared with $\sigma_{\rm{stat}}$ at $Q_{\beta\beta}$.
Because $\sigma_{\rm PedErr}$ is proportional to the integration interval, signal integration is not appropriate for calculating the energy. 
An alternative method for CANDLES~III is proposed to obtain the energy information with the least effect of baseline fluctuations on the energy.

\section{Photon counting in CANDLES III}
\label{sec:PC}
\subsection{DAQ modification for photon counting}
\label{ssec:setup}
The unavoidable fluctuation induced by the pedestal uncertainty is accumulated in the signal integration.
The photon counting method is widely used in scintillator experiments to reduce baseline fluctuations.
The sum waveform of the CaF$_2$ signal is formed by up to several thousands of p.e. signals, and each PMT waveform contains less $N_{\rm p.e.}$.
The overlap of 1~p.e. signals results in inefficiency in photon counting, as mentioned in \ref{ssec:overlap}; hence, we should count $N_{\rm p.e.}$ in each PMT of CANDLES.
Currently, the first 768~ns of each PMT waveform is digitized by an FADC every 2~ns and recorded as 8-bit data; thereafter,  digitized values in every 64~ns are summed and recorded as 16-bit data \cite{Khai2019}. 
The waveform interval of each FADC is 8960~ns, and the data size is 640~B/FADC/event \cite{Khai2019}.
The 1~p.e. width of each PMT is less than 50~ns; thus, the shape of the 1~p.e. signal is difficult to observe if the signal rises after 768~ns.
The DAQ software is modified to record the first 4088~ns of the waveform at a speed of 2~ns/sample, and records 128~ns by summing the digitized values for every 64~ns. 
After DAQ modification, the waveform interval is reduced to 4216~ns, and the data size is increased to 2048~B/FADC/event, which is the buffer size limit in each FADC.
The different waveforms obtained by normal DAQ and modified DAQ can be seen in Figure~\ref{fig:waveform} and Figure~\ref{fig:waveform2}~(top), respectively.
Based on data from a previous study \cite{Khai2019}, the readout time per event for photon counting measurement is 20~ms/event, which is twice as long as the current readout time in the physics run of CANDLES \cite{Khai2019}.
Owing to the development of the DAQ with eight buffers acting as derandomizers \cite{Khai2019} in each FADC, the data-taking efficiency with the increased data size is still almost 100$\%$.

\subsection{Overlap of single p.e. signals}
\label{ssec:overlap}
The threshold for photon counting should not be set too low to avoid  baseline noise, or too high to avoid losing p.e. in counting.
In this study, the threshold for each PMT is ${\rm Pedestal}-\left(\mu_{\rm p}-2\sigma_{\rm p}\right)$, where $\mu_{\rm p}$ and $\sigma_{\rm p}$ are the mean and standard deviation, respectively, of 1~p.e. pulse height distribution of the corresponding PMT.
This threshold is set for every PMT because it provides a good separation between the baseline and non-baseline signals.
For every time bin in 4088~ns of the waveform, if the signal crosses over the threshold, it is counted as 1~p.e.
With simple photon counting, a multi-p.e. signal is counted as a single p.e. signal, which leads to missing p.e. in counting.
\par
\begin{figure}[h!]
	\centering
	\includegraphics[width=0.8\linewidth, clip, trim={0mm 0mm 0mm 0mm}]{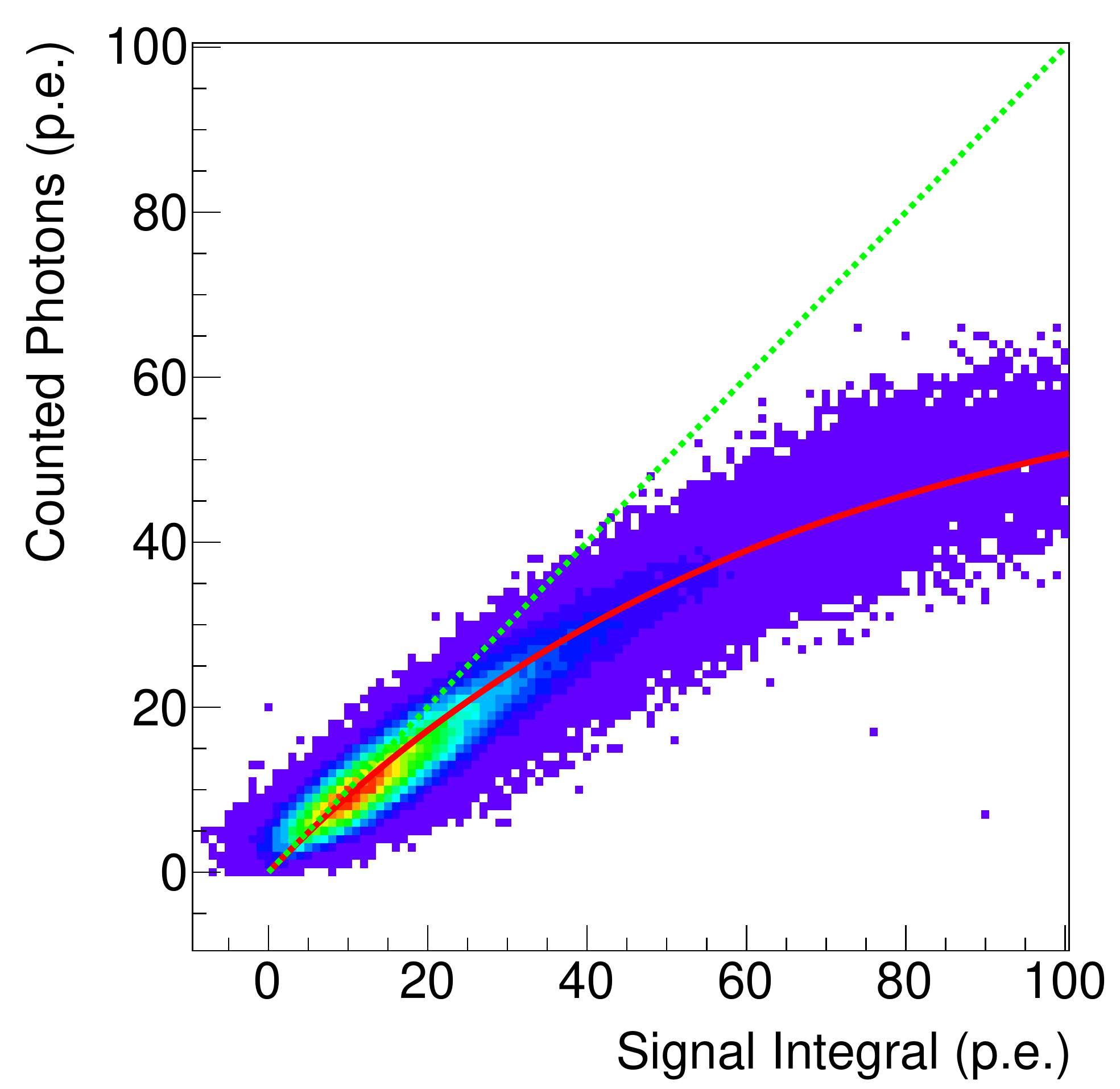}\hfill
	\caption{Counting efficiency in a PMT.
		The green dashed line is the expected linearity, and the solid red line is the fitting function.
	}
	\label{fig:check1peetc010-188photons-int10area}
\end{figure}

If the time interval between the two 1~p.e. signals is too short, it is impossible to distinguish them in photon counting. 
We use a simple mathematical model to estimate the number of counted p.e. from the number of true p.e.
We assume that there are two adjacent 1~p.e. signals in one PMT, named ``A" and ``B", respectively, and define “signal interval” ($w_{\rm s}$) as the shortest interval for distinguishing these two signals. 
The $w_{\rm s}$ value is related to the width of the 1~p.e. signal, which corresponds to the rise time and fall time of the PMT, and the threshold set for photon counting.
To avoid missing signal B in counting, there should be no signal in the $w_{\rm s}$~ns preceding signal A. 
The probability of counting signal B is the probability of no signal in the $w_{\rm{s}}$~ns preceding signal A: ${\rm e}^{-\mu(t)w_s}$, where $\mu(t)$ is defined in equation \ref{eq:waveform}. 
The number of counted p.e. ($N_{\rm c}$) is
\begin{equation}
	\begin{split}
		dN_{\rm c} &= \mu(t){\rm e}^{-\mu(t)w_{\rm s}}dt\\
		\Rightarrow N_{\rm c} &= \int_{0}^{\infty}dN_c
		= \frac{\tau}{w_{\rm s}}\left(1-{\rm e}^{-N_{\rm p.e.}w_{\rm s}/\tau}\right).
	\end{split}
	\label{eq:CountPE}
\end{equation}
The counting efficiency in a PMT is evaluated in Figure~\ref{fig:check1peetc010-188photons-int10area} by checking the correlation of the counted p.e. and the signal integration.
The green dashed line indicates the expected 100$\%$ counting efficiency, and the solid red line is the fitting function using equation \ref{eq:CountPE}.
At $Q_{\beta\beta}$($^{48}$Ca), the number of p.e. is approximately 63 p.e./PMT, whereas the number of counted p.e. in this PMT is approximately 40 p.e.
Two histograms constructed using signal integration in 4000~ns and photon counting in 4000~ns are plotted in black and magenta, respectively, in Figure~\ref{fig:energyhistps2sigma}.
It is very clear that the histogram using 4000~ns photon counting has a worse energy resolution because many p.e. were missed during the counting.
\par
\begin{figure}[h!]
	\centering
	\includegraphics[width=0.8\linewidth, clip, trim={0mm 0mm 0mm 0mm}]{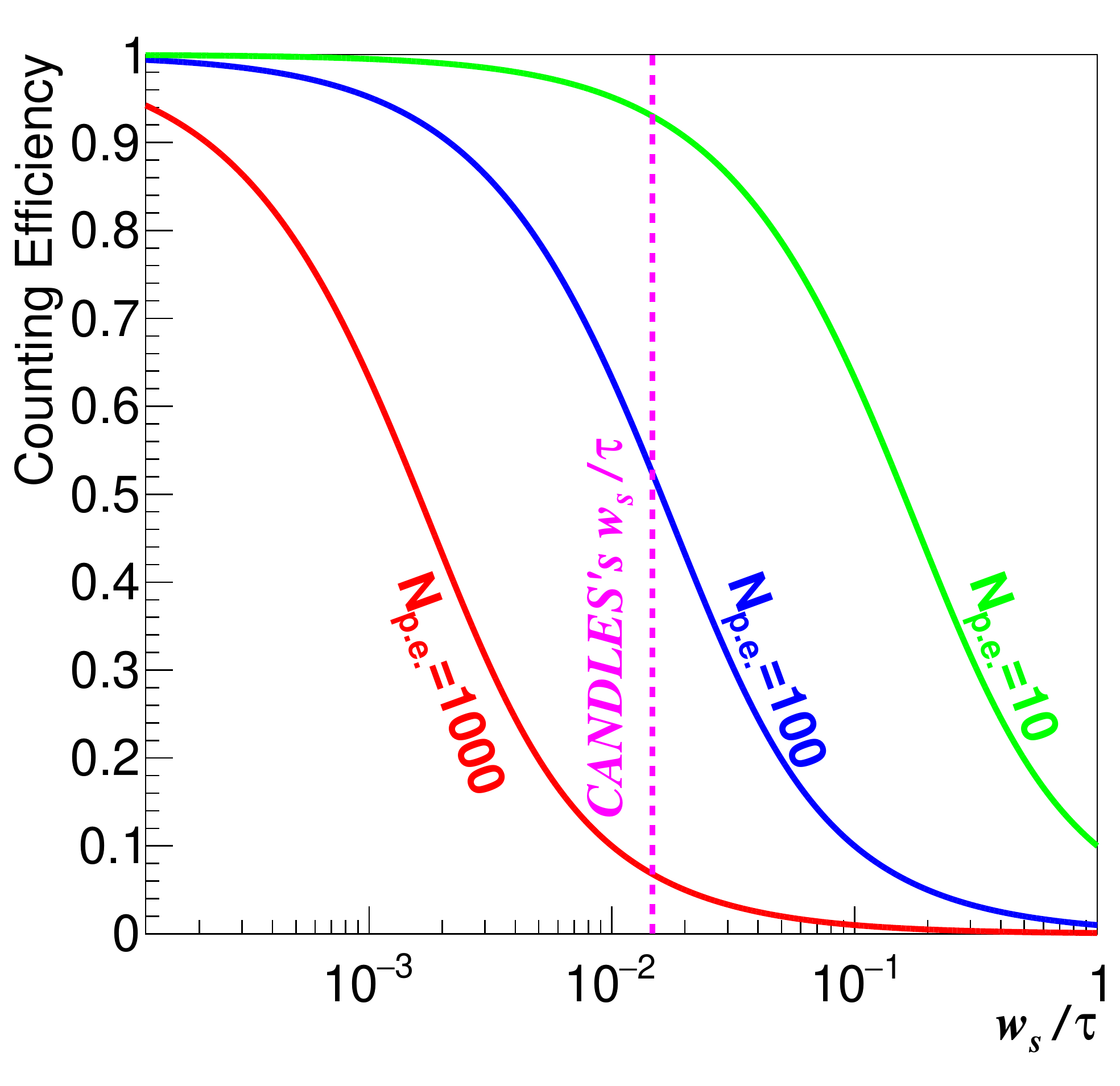}\hfill
	\caption{Estimated counting efficiency in one PMT as a function of $w_{\rm s}/\tau$.
		Different $N_{\rm p.e.}$ values of 10, 100, and 1000 are plotted in green, blue and red, respectively.
		The dashed magenta line indicates the average $w_{\rm s}/\tau$ value of 62 PMTs.
	}
	\label{fig:counteff}
\end{figure}

Counting efficiency is essential in photon counting, and it is affected by the number of incident p.e., signal interval, and decay constant.
Figure~{\ref{fig:counteff}} shows the estimated counting efficiency ($N_{\rm c}/N_{\rm p.e.}$) for one PMT as a function of $w_{\rm s}/\tau$ by using equation {\ref{eq:CountPE}}.
This estimation was separately performed using different $N_{\rm p.e.}$ values of 10, 100, and 1000 p.e.
The dashed magenta line marks the average $w_{\rm s}/\tau$ value obtained for 62~PMTs in CANDLES.
According to Figure~{\ref{fig:counteff}}, the counting efficiency can be improved by reducing the number of incident p.e. obtained in a PMT; this can be achieved by introducing many PMTs surrounding the scintillator or using the detector to measure low-energy regions.
Another way to improve the counting efficiency is reducing $w_{\rm s}/\tau$, by using a scintillator material with a longer decay constant, or selecting the PMTs that provide a shorter 1~p.e. signal width.

\subsection{Partial photon counting}
\label{ssec:PPC}
\begin{figure}[h!]
	\includegraphics[width=1.0\linewidth]{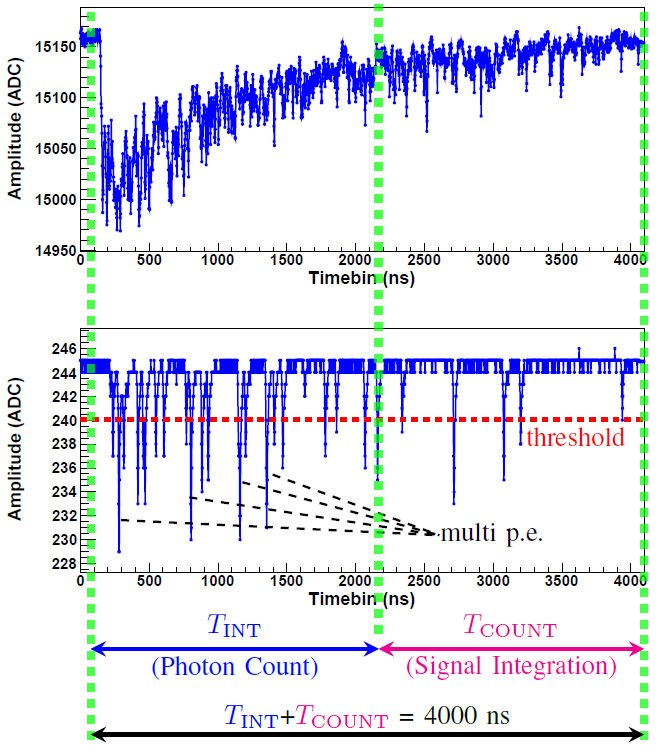}
	\caption{Sum waveform of 62 PMTs (top) and waveform of one PMT (bottom).
		The sum waveform contains many photoelectrons, but there are fewer photoelectrons in one PMT.
		In PPC, the 4000~ns waveform of every PMT is divided into two areas: the prompt region for signal integration, and the tail region for photon counting.
	}
	\label{fig:waveform2}
\end{figure}
\begin{figure}[h!]
	\centering
	\includegraphics[width=0.99\linewidth,page=1]{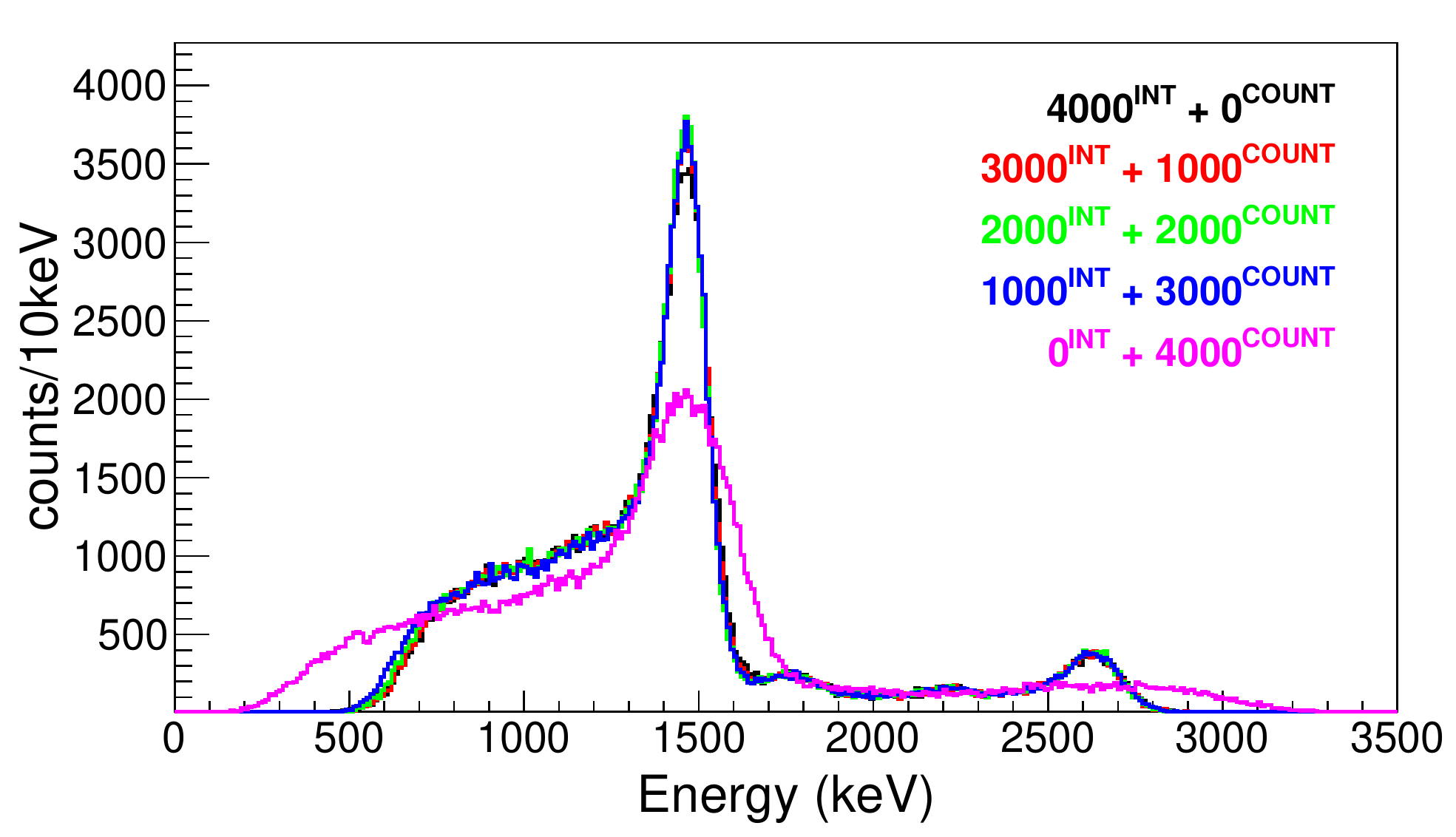}
	\caption{
		Energy histograms constructed using the PPC method after calibration.
		Histograms using different mixtures of integration and photon-counting intervals are shown.
	}
	\label{fig:energyhistps2sigma}
\end{figure}

The severe baseline fluctuations are accumulated in the 4000~ns integration interval (Section~\ref{sec:ErrCharge}); therefore, it is encouraged to use photon counting instead of signal integration in CANDLES~III. 
However, the spectrum constructed using 4000~ns photon counting has a poor energy resolution because the overlap of 1~p.e. signals leads to missing p.e. in counting (Section~\ref{ssec:overlap}).
In this section, we introduce a method named ``partial photon counting" (PPC) to reduce the baseline fluctuation with the fewest possible missed p.e. in photon counting. 
The multi-p.e. signals are found predominantly near the rising edge of the CaF$_2$ waveform.
Thus, each PMT waveform is divided into two regions: in the prompt region near the rising edge, where many multi-p.e. signals are found, signal integration is used to avoid missing p.e.; and in the latter region near the tail, where only a few multi-p.e. signals are found, photon counting is performed to reduce the baseline fluctuation. 
The sum of the integration and photon-counting intervals is fixed at 4000~ns.
From summing the waveforms of 62~PMTs, we can obtain the energy of the CaF$_2$ signal.
The details of the PPC method are provided in Figure~\ref{fig:waveform2}.
Some preliminary results of improved energy resolutions obtained with the PPC method and a Gaussian-plus-exponential fitting function are introduced in reference \cite{9059670}.
Energy histograms are constructed with different mixtures of integration and photon-counting intervals to evaluate the performance of PPC.
Several histograms constructed using the PPC method are shown in Figure~\ref{fig:energyhistps2sigma} with different mixtures including 4000~ns (integration), 3000~ns (integration) + 1000~ns (counting), 2000~ns (integration) + 2000~ns (counting), 1000~ns (integration) + 3000~ns (counting), and 4000~ns (counting), plotted in black, red, green, blue, and magenta, respectively.
Because the counting efficiency is not 100$\%$, the energy peaks are left-shifted when the photon-counting interval is increased. 
Therefore, all energy histograms in Figure~\ref{fig:energyhistps2sigma} are calibrated using $\gamma$-peaks of $^{40}$K and $^{208}$Tl in this research for ease of further analysis.
\par
\begin{figure}[t!]
	\centering
	\includegraphics[width=0.99\linewidth,page=1,clip,trim={0mm 0mm 0mm 0mm}]{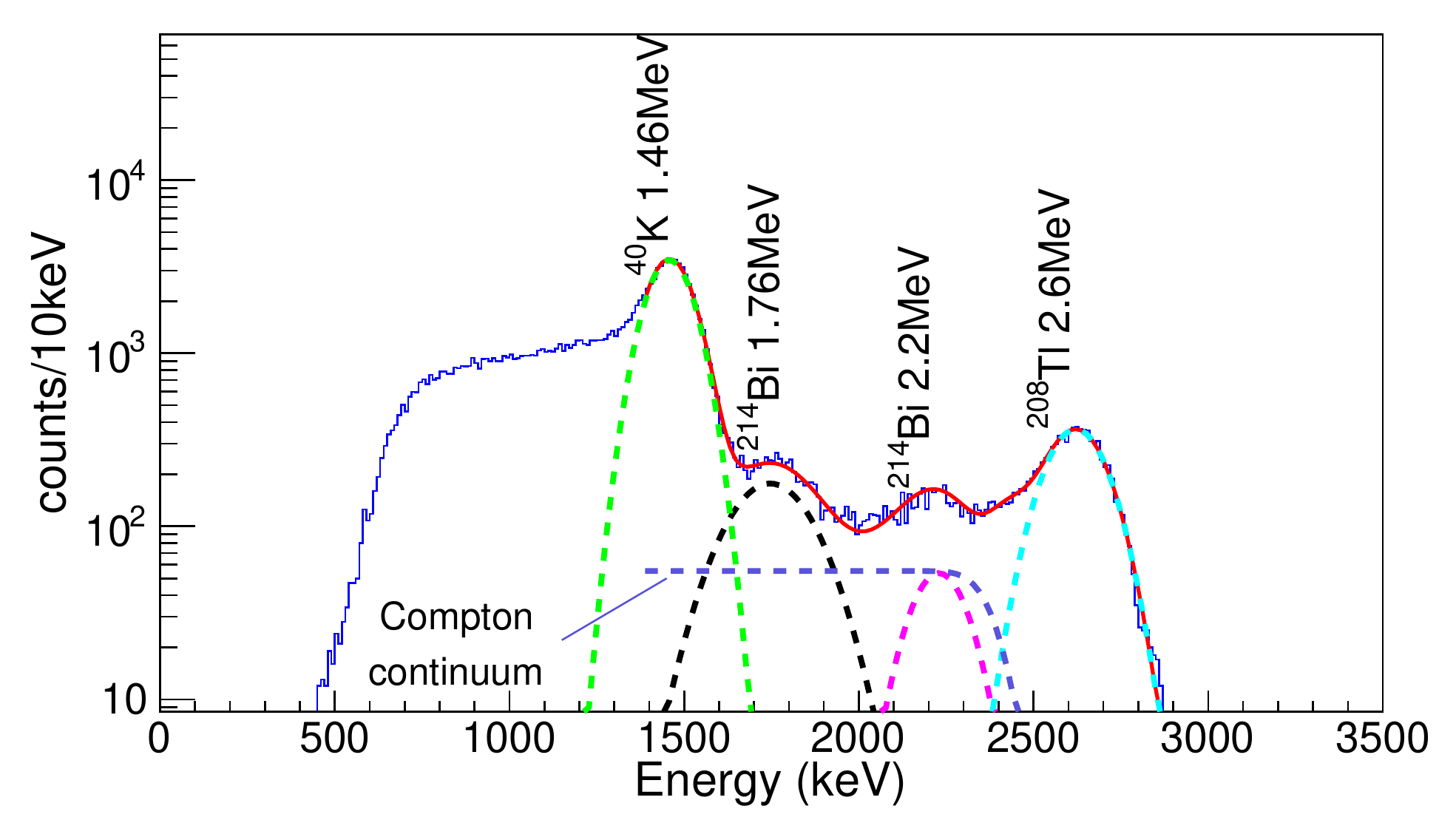}
	\caption{
		Fitting function (solid red line) applied to every energy spectrum to obtain the energy resolutions of the $^{40}$K and $^{208}$Tl peaks.
		The fitting function contains Gaussian peaks ($\gamma$-peaks of $^{208}$Tl, $^{214}$Bi, and $^{40}$K) and an error function representing the Compton continuum.
	}
	\label{fig:fithistps2}
\end{figure}
\begin{figure}[b!]
	\centering
	\includegraphics[width=0.99\linewidth,page=1]{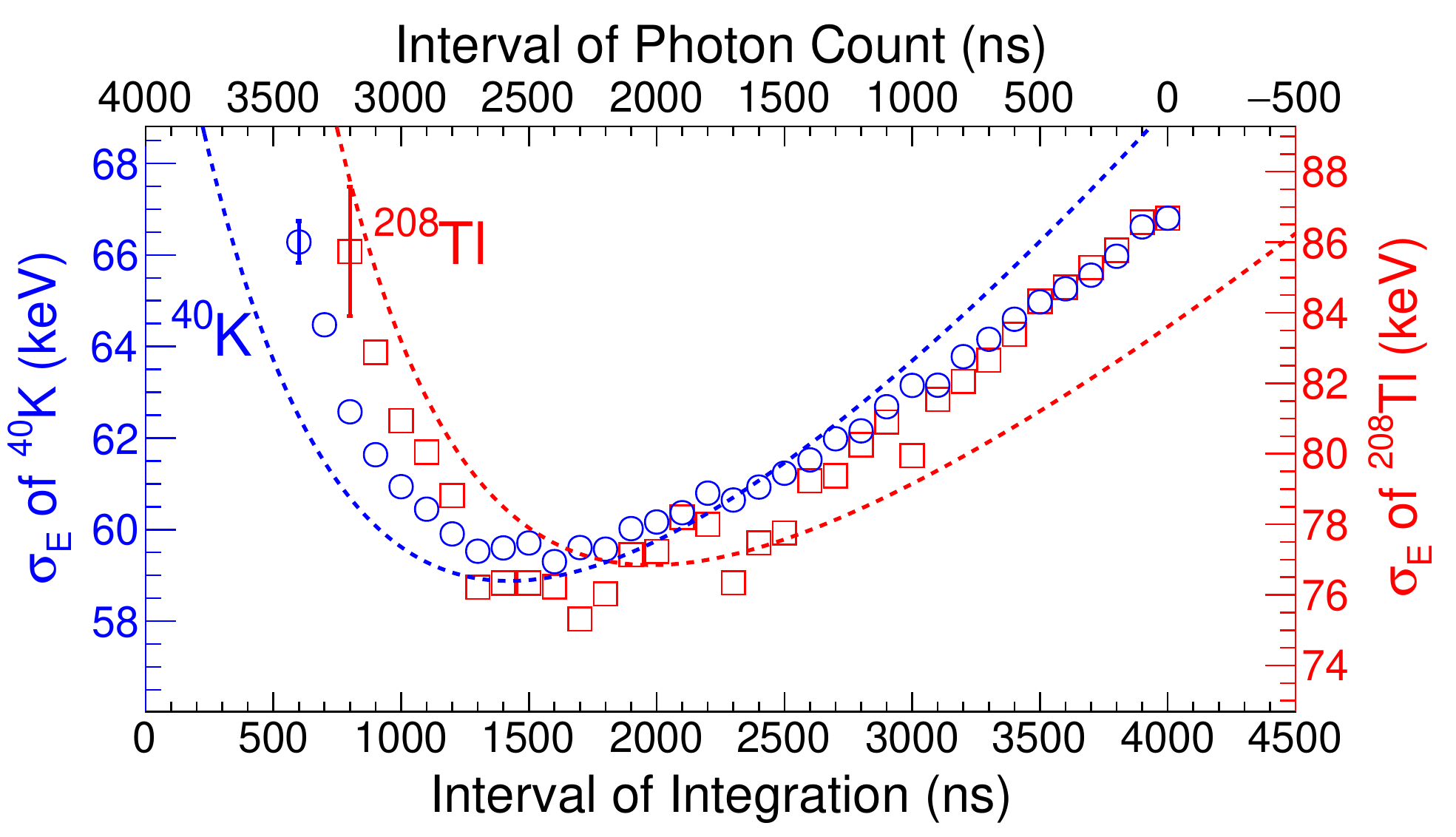}
	\caption{Obtained energy resolution at $^{40}$K (blue circles) and $^{208}$Tl (red squares) peaks as a function of integration interval.
		The estimated resolutions at the $^{40}$K and $^{208}$Tl peaks are plotted with blue and red dashed lines, respectively. The lines are plotted using equation \ref{eq:SumSigma}. 
	}
	\label{fig:resolution}
\end{figure}

Each energy histogram obtained in PPC is fitted with a function, which is a sum of Gaussian functions, representing the $\gamma$-peaks emitted from $^{40}$K (1.46~MeV {\cite{NuclData}}), $^{214}$Bi (1.76~MeV and 2.2~MeV {\cite{NuclData}}), and $^{208}$Tl (2.6~MeV {\cite{NuclData}}), and an error function representing the Compton continuum.
Figure~\ref{fig:fithistps2} shows the energy spectrum constructed with a 4000~ns integration interval with the fitting function.
The energy resolutions at $^{40}$K and $^{208}$Tl peaks in each histogram are checked.
To evaluate the performance of the PPC method, the resolutions are plotted as a function of the integration interval in Figure~\ref{fig:resolution}. 
The estimated resolutions at the $^{40}$K and $^{208}$Tl peaks are plotted using blue and red dashed lines, respectively.
The estimated resolutions are calculated using equation \ref{eq:SumSigma}, which is the root sum square of all fluctuations examined in this study. 
The differences between the experimental and estimated values at both energy peaks may be caused by imperfect estimation of the statistical fluctuation, which is discussed in the next paragraph, and the assumption of residual fluctuations, which is discussed in \ref{ssec:fluc_summary}.
The differences between the experimental and estimated values are small (less than 4~keV) at both energy peaks.
For each $\gamma$-peak, the errors of adjacent data points are correlated; therefore, only one error bar at one integration interval is shown. 
Owing to the small number of events at the $^{208}$Tl peak, we found the deviation of the obtained resolutions at this energy peak as well as a wide error bar.
The energy resolutions at the two energies are improved owing to the reduction of baseline fluctuation when the integration interval is decreased.
Near the rising edge, the energy resolutions degrade because more p.e. are missed, due to the high overlapping probability of 1~p.e. signals.
Because more p.e. are obtained at the $^{208}$Tl peak compared to the $^{40}$K peak, the overlapping probability of 1~p.e. signals at the $^{208}$Tl peak is higher compared to that at the $^{40}$K peak.
The integration interval to obtain the best energy resolution at the $^{208}$Tl peak should be longer than that at the $^{40}$K peak to limit the p.e. missed in counting. 
From the results, we obtain the energy deviation ($\sigma_{\rm{E}}$) improved from 66.8 to 59.3~keV, or energy resolution ($\sigma_{\rm{E}}$/E) improved from 4.5$\%$ to 4.0$\%$, at the $^{40}$K peak when reducing the integration interval from 4000~to~1600~ns; and $\sigma_{\rm{E}}$ improved from 86.7 to 75.3~keV, or $\sigma_{\rm{E}}$/E improved from 3.3$\%$ to 2.9$\%$ at the $^{208}$Tl peak when reducing the integration interval from 4000 to 1700~ns.
With the results at the $^{40}$K and $^{208}$Tl peaks, it is expected to obtain an improved resolution at $Q_{\beta\beta}$($^{48}$Ca).
\par
The overlap of 1~p.e. signals degrades the energy resolution because of the increment in the statistical fluctuation. 
The statistical fluctuation in PPC is estimated using the following mathematical model. 
The regions in PPC include the prompt region for integration and the latter region for photon counting.
With the predefined parameters in equation \ref{eq:waveform}, the number of p.e. in the prompt region ($N_{\rm P}$) is 
\begin{equation}
	\ N_{\rm P} = \int_{0}^{T_{\rm{INT}}}\mu(t)dt = N_{\rm p.e.}\left(1-{\rm e}^{-T_{\rm{INT}}/\tau}\right).
	\label{eq:StatPPC_NP}
\end{equation} 
The number of p.e. in the latter region is $N_{\rm D} = N_{\rm p.e.}{\rm e}^{-T_{\rm{INT}}/\tau}$, and the number of counted p.e. in the latter region of a PMT waveform, $M_{\rm D}$, is calculated using equation \ref{eq:CountPE}:
\begin{equation}
	M_{\rm D} = \frac{\tau}{w_{\rm s}}\left(1-{\rm e}^{-N_{\rm D}w_{\rm s}/\tau}\right).
	\label{eq:StatPPC_MD}
\end{equation}
The obtained number of p.e. ($N_{\rm{T}}$) is the sum of $N_{\rm{P}}$ and $M_{\rm{D}}$.
Because $N_{\rm{T}}$ is normalized to $N_{\rm p.e.}$ in the analysis, the statistical fluctuation after calibration is
\begin{equation}
	\Delta N_{\rm T} = \sqrt{\Delta N_{\rm P}^2 + \Delta M_{\rm D}^2 \left(\frac{dN_{\rm D}}{dM_{\rm D}}\right)^2}.
	\label{eq:StatPPC_DeltaNT}
\end{equation}
Applying equation \ref{eq:StatPPC_DeltaNT} with different mixtures of integration and photon-counting intervals, the statistical fluctuations in the PPC method at the $^{40}$K and $^{208}$Tl peaks are estimated.
The statistical fluctuations at these two peaks are plotted as functions of the integration interval with the two solid black lines in Figure \ref{fig:resolutionk40tl208_2}.

\section{Results and discussion}
\label{sec:results}
\subsection{Discussion of the fluctuations}
\label{ssec:fluc_summary}
\begin{figure}[h!]
	\centering
	\includegraphics[width=0.99\linewidth,page=1,clip,trim={0mm 3mm 0mm 0mm}]{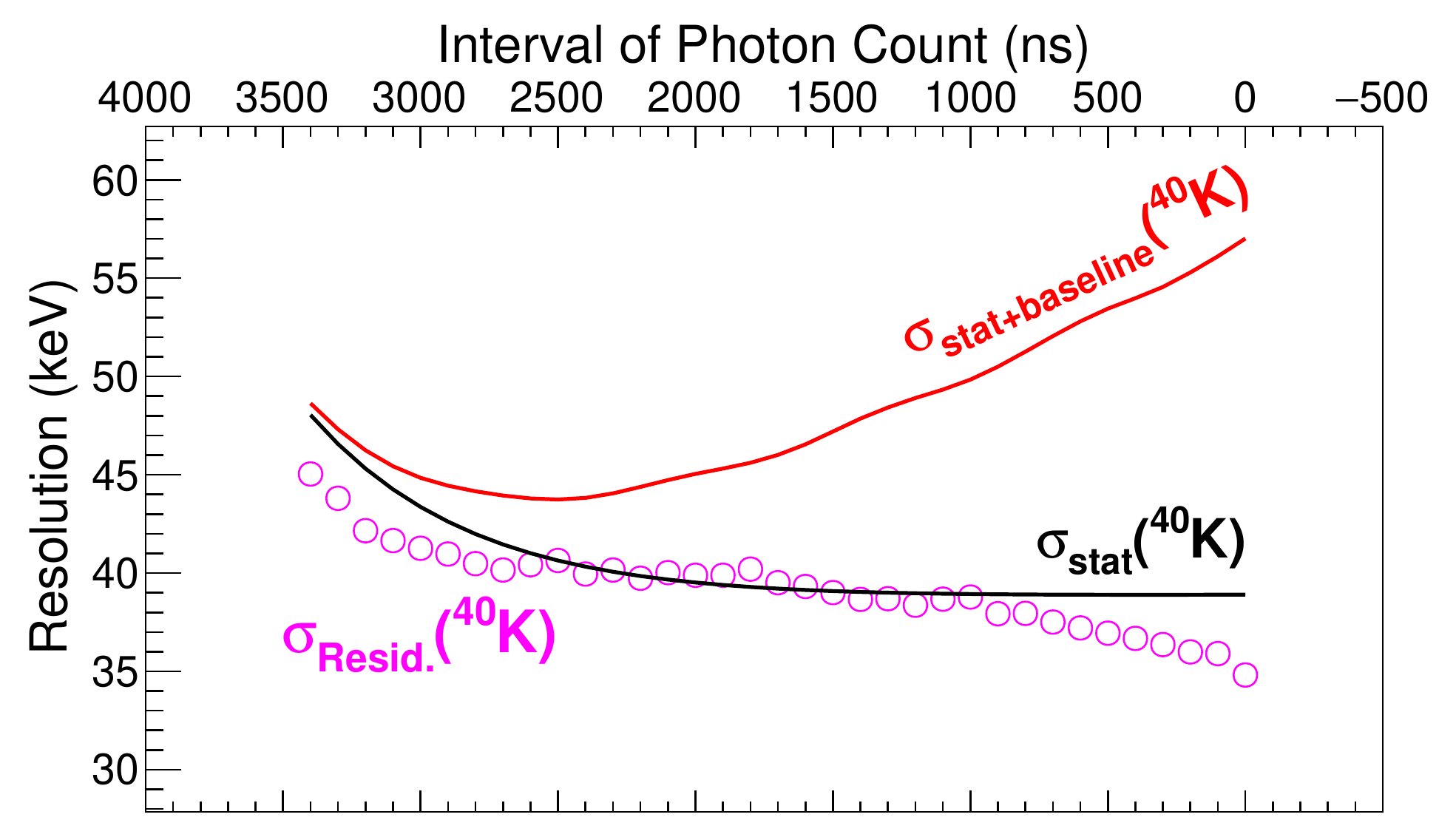}\hfill
	\includegraphics[width=0.99\linewidth,page=1,clip,trim={0mm 0mm 0mm 0mm}]{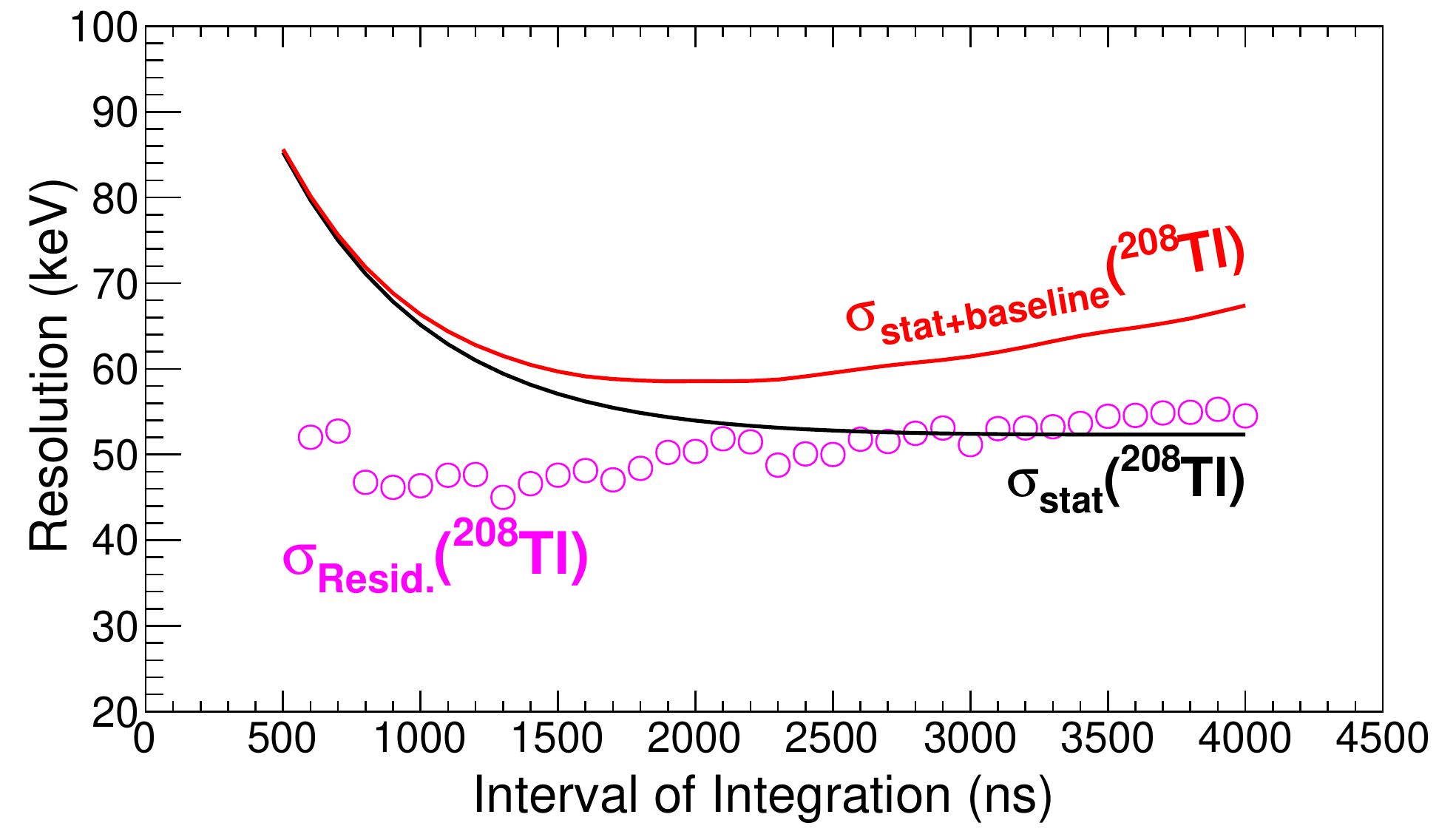}
	\caption{Statistical fluctuations, statistical+baseline fluctuations, and residual fluctuations at $^{40}$K and $^{208}$Tl $\gamma$-peaks are plotted as a function of the integration interval.}
	\label{fig:resolutionk40tl208_2}
\end{figure}
In Figure \ref{fig:resolutionk40tl208_2}, we summarize the fluctuations at the $^{40}$K and $^{208}$Tl $\gamma$-peaks and plot them as functions of integration interval. 
The statistical fluctuations estimated by equation \ref{eq:StatPPC_DeltaNT} are shown as solid black lines.
Extending the photon-counting interval increases the number of 1~p.e. signals overlapping each other; hence, those p.e. are missed in counting, and the statistical fluctuation becomes worse.
The root sum squares of the statistical and baseline fluctuations, $\rm{\sqrt{\sigma^2_{stat}+\sigma^2_{baseline}}}$, at the two energy peaks are plotted as solid red lines.
The baseline fluctuations include the pedestal uncertainty, $\rm{\sigma_{PedErr}}$, and the DE fluctuation, $\rm{\sigma_{\rm{DE}}}$. 
Because the fluctuations induced by DC, $\rm{\sigma_{DC}}$, and sinusoidal noise, $\rm{\sigma_{noise}}$, are negligibly small, they are not accounted for in the baseline fluctuations.
Additionally, photon counting cannot remove the DC 1~p.e. signals in the CaF$_2$ waveform; therefore, the DC fluctuation cannot be reduced by the PPC method.
In addition to the baseline fluctuation and estimated statistical fluctuation, we study the residual fluctuation as a function of integration interval.
The residual fluctuations, $\rm{\sigma_{Resid.}=\sqrt{\sigma^2_{E}-\sigma^2_{baseline}-\sigma^2_{stat}}}$, of the two energy peaks are plotted as magenta points.
Despite some small uncertainties, the ${\rm{\sigma_{Resid.}}}$ at either the $^{40}$K or $^{208}$Tl peak is almost unchanged within the 600--4000~ns integration interval.
Therefore, the tendency of resolution enhancement using the PPC method can be explained by the reduction of the baseline fluctuation.
Because the average ${\rm{\sigma_{Resid.}}}$ within the integration interval of 600--4000~ns  for each energy peak is close to the statistical fluctuation at 4000 ns, ${\rm{\sigma_{Resid.}}}$ can be assumed to be dependent on the energy and independent of the integration interval.
In this paper, the source of the residual fluctuations is not discussed.
\par
\begin{figure}[h!]
	\centering
	\includegraphics[width=0.99\linewidth,clip,trim={0mm 0mm 0mm 0mm}]{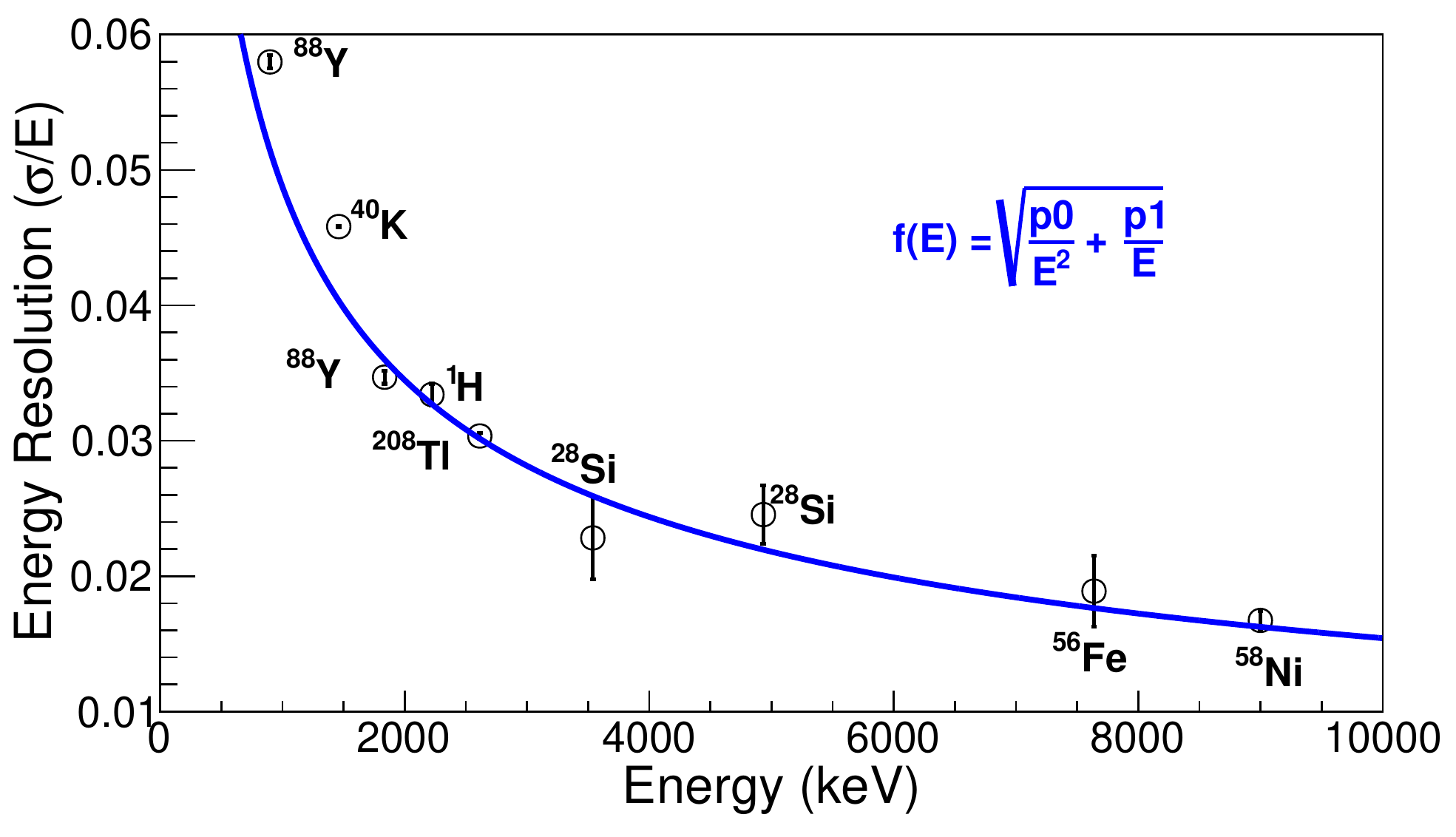}
	\caption{Application of the fitting function in equation \ref{eq:Res} on the obtained resolutions in Run010 of CANDLES.
		The data points are referred from \cite{NIMA-IIDA}.}
	\label{fig:resolutionrun010}
\end{figure}

In summary, the energy resolution can be expressed using the following function:
\begin{equation}
	\rm{\sigma_{E}} = \rm{\sqrt{\sigma^2_{stat}+\sigma^2_{DE}+\sigma^2_{Resid.}+\sigma^2_{PedErr}}},
	\label{eq:SumSigma}
\end{equation}
where $\rm{\sigma_{stat}}$ is the statistical fluctuation, $\rm{\sigma_{DE}}$ is the DE fluctuation, $\rm{\sigma_{Resid.}}$ is the residual fluctuation, and $\rm{\sigma_{PedErr}}$ is the accumulated fluctuation induced by pedestal uncertainty.
The estimated DE corresponds to $N_{\rm{p.e.}}$; thus, the $\rm{\sigma_{DE}}$ fluctuation depends on $\rm{\sqrt{E}}$.
From the above discussion, the fluctuations studied in this research can be categorized into two groups: energy-dependent fluctuations ($\rm{\sigma_{stat}}$, $\rm{\sigma_{DE}}$, and $\rm{\sigma_{Resid.}}$), which are proportional to $\rm{\sqrt{E}}$, and energy-independent fluctuation ($\rm{\sigma_{PedErr}}$). 
We assume that the energy resolution can be estimated by the following fitting function:
\begin{equation}
	\frac{\sigma_{\rm E}}{\rm E} = \sqrt{\frac{p_0}{\rm E^2}+\frac{p_1}{\rm E}}.
	\label{eq:Res}
\end{equation}
The goodness of the above fitting function is evaluated by applying it to the energy resolutions taken from \cite{NIMA-IIDA}, obtained at different energies.
In Figure \ref{fig:resolutionrun010}, the energy peaks consist of $\gamma$-peaks from radioisotopes ($^{40}$K, $^{208}$Tl, and $^{88}$Y), and $\gamma$-peaks from ($n$, $\gamma$) reactions on $^{1}$H, $^{28}$Si, $^{56}$Fe, and $^{58}$Ni.
The details of ($n$, $\gamma$) calibration for the CANDLES III detector and the emitted $\gamma$-peaks can be found in \cite{NIMA-IIDA}.

\subsection{Estimating the improved sensitivity of CANDLES III}
\label{ssec:sensitivity}
\begin{figure}[h!]
	\centering
	\includegraphics[width=0.99\linewidth,clip,trim={0mm 0mm 0mm 0mm}]{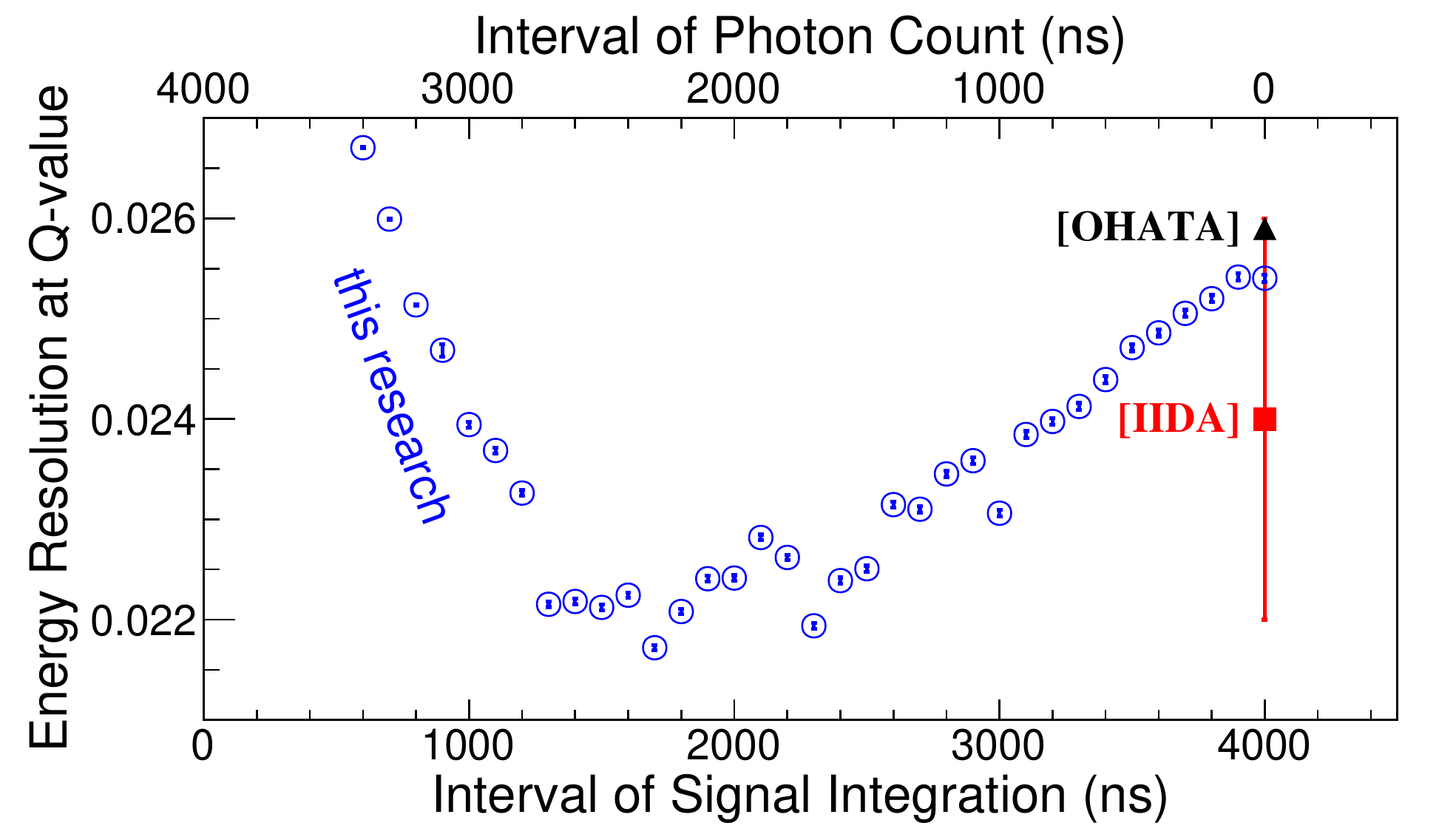}
	\caption{Estimated energy resolutions at the Q-value using the PPC method.
		The blue circles represent the $\sigma_{\rm{E}}$($Q_{\beta\beta}$) estimated in this study.
		The black triangle and red square indicate the $\sigma_{\rm{E}}$($Q_{\beta\beta}$) obtained in previous studies by Ohata \cite{Ohata_DThesis} and Iida \cite{NIMA-IIDA}, respectively.}
	\label{fig:testca48ps2}
\end{figure}

For each mixture of integration and photon-counting intervals, equation \ref{eq:Res} is used to fit the energy resolutions at the $^{40}$K and $^{208}$Tl peaks. The energy resolution at $Q_{\beta\beta}$($^{48}$Ca) is then extrapolated. 
The estimated results are plotted as a function of integration interval with blue circles in Figure \ref{fig:testca48ps2}.
The uncertainty of the blue circles at the 1200--2600~ns integration interval, shown in Figure \ref{fig:testca48ps2}, is due to the uncertainty of the resolutions obtained at the $^{40}$K and $^{208}$Tl peaks plotted in Figure \ref{fig:resolution}.
The energy resolutions at $Q_{\beta\beta}$($^{48}$Ca), reported by Ohata \cite{Ohata_DThesis} and Iida \cite{NIMA-IIDA}, are obtained with the 4000~ns integration method, and plotted with black triangles and red squares, respectively.
With an integration interval of 4000~ns, there is an agreement between the estimated resolution at $Q_{\beta\beta}$($^{48}$Ca) in this study and those obtained in previous studies. 
From the figure, the energy resolution at $Q_{\beta\beta}$($^{48}$Ca) can be improved from 2.6$\%$ to 2.2$\%$ by using the PPC method when the integration interval is reduced from~4000~to~2300~ns. 
\par
The sensitivity of $T_{1/2}^{0\nu}$ is related to the background rate and energy resolution.
In the current CANDLES III, the 2$\nu\beta\beta$ background is not dominant compared to the natural background, and the sensitivity of $T_{1/2}^{0\nu}$ is proportional to the inverse square root of the energy resolution at $Q_{\beta\beta}$ \cite{Rev0nbb}.
The energy resolution at $Q_{\beta\beta}$($^{48}$Ca) is estimated to be improved from 2.6$\%$, by using a 4000~ns integration, to approximately 2.2$\%$, by using PPC. 
Therefore, using the PPC method can improve the sensitivity of the CANDLES III detector for $T_{1/2}^{0\nu}$ of $^{48}$Ca by $\sqrt{2.6\%/2.2\%}=1.09$ times.

\section{Summary}
\label{sec:summary}
CANDLES aims to observe the 0$\nu\beta\beta$ of $^{48}$Ca using CaF$_2$ crystals. 
The 2$\nu\beta\beta$ is an irremovable background in CANDLES, and it can be a severe background in our future ton-scale detector with CaF$_2$(un-doped, $^{48}$Ca-enriched) crystals.
The energy resolution is essential to reduce the 2$\nu\beta\beta$ as well as other background events.
At $Q_{\beta\beta}$($^{48}$Ca), the energy resolution, 2.6$\%$, is larger than the ideal statistical fluctuation of the number of p.e., 1.6$\%$, and there are other fluctuations that worsen the energy resolution.
The baseline fluctuations are accumulated in the 4000~ns signal integration, which is used to calculate the energy of CaF$_2$. 
In this study, the baseline fluctuations, including dark current, sinusoidal noise, digitization error, and pedestal uncertainty are investigated.
The fluctuation induced by pedestal uncertainty is found to be the most severe fluctuation, 1$\%$ at $Q_{\beta\beta}$($^{48}$Ca).
\par
Photon counting is useful for removing the baseline fluctuations, but it results in missing p.e. in counting for each PMT and, consequently, a worse energy resolution.
We introduce a waveform analysis method named ``partial photon counting," in which the signal integration is carried out in the prompt region and the photon counting is carried out in the tail region, to improve the energy resolution.
Using this method, we obtain an improvement in the energy resolutions at $\gamma$-peaks of $^{40}$K and $^{208}$Tl, and the energy resolution at $Q_{\beta\beta}$ is estimated to be improved to 2.2$\%$.
With this improvement, we expect the sensitivity of $T_{1/2}^{0\nu}$ of $^{48}$Ca to be improved by 1.09 times using the same detector status as that reported in \cite{Ohata_DThesis}.

\section{Acknowledgment}
The authors thank Mr.~Keita Mizukoshi from Kobe University and Dr.~Michael Moser from Siemens~Healthineers for their helpful discussions and comments.

\bibliographystyle{ieeetr}
\bibliography{reference}

\end{document}